\documentclass[useAMS,usenatbib]{mnras}
\usepackage[utf8]{inputenc}
\usepackage{amssymb,amsmath}
\usepackage{graphicx}
\usepackage[dvipsnames]{xcolor}
\hypersetup{colorlinks=true,allcolors=teal}

\newcommand{\be}{\begin{equation}}
\newcommand{\ee}{\end{equation}}
\def\bear#1\ear{\begin{align}#1\end{align}}
\newcommand{\nline}{\notag \\}
\newcommand{\f}{\frac}
\newcommand{\de}{\mathrm{d}}

\newcommand{\e}{\mathrm{e}}

\newcommand{\Msun}{\mathrm{M}_{\odot}}

\newcommand{\eqn}[1]{equation (\ref{#1})}

\newcommand{\secn}[1]{Section \ref{#1}}
\newcommand{\fig}[1]{Fig. \ref{#1}}

\title[Ly$\alpha$ opacity fluctuations using fast methods]{Studying the Lyman-$\alpha$ optical depth fluctuations at $z \sim 5.5$ using fast semi-numerical methods} 
\author[Choudhury et al]{
T. Roy Choudhury$^1$\thanks{E-mail: tirth@ncra.tifr.res.in},
Aseem Paranjape$^2$\thanks{Email: aseem@iucaa.in}
and
Sarah E. I. Bosman$^3$
\\
$^1$National Centre for Radio Astrophysics, TIFR, Post Bag 3, Ganeshkhind, Pune 411007, India\\
$^2$Inter-University Center for Astronomy \& Astrophysics, Post Bag 4, Ganeshkhind, Pune 411007, India\\
$^3$Department of Physics and Astronomy, University College London, London, UK
}
\begin{document} 
 
\date{} 

\maketitle

\begin{abstract}
We present a computationally efficient and fast semi-numerical technique for simulating the Lyman-$\alpha$ (Ly$\alpha$) absorption optical depth in presence of neutral hydrogen ``islands'' left over from reionization at redshifts $5 \lesssim z \lesssim 6$. 
The main inputs to the analysis are (i) a semi-numerical photon-conserving model of ionized regions during reionization (named \texttt{SCRIPT}) along with a prescription for simulating the shadowing by neutral islands and (ii) the fluctuating Gunn-Peterson approximation to model the Ly$\alpha$ absorption.
Our model is then used for simulating the large-scale fluctuations in the effective optical depth as observed along sight lines towards high$-z$ quasars.
Our model is fully described by five parameters.
By setting two of them to default values and varying the other three, we obtain the constraints on reionization history at $5 \lesssim z \lesssim 6$ as allowed by the data.
We confirm that reionization is \emph{not} complete before $z \sim 5.6$ at $\gtrsim 2\sigma$ confidence, with the exact confidence limits depending on how the non-detections of the flux in the data are treated.
We also confirm that the completion of reionization can be as late as $z \sim 5.2$.
With further improvements in the model and with more sight lines at $z \sim 6$, we can take advantage of the computational efficiency of our analysis to obtain more stringent constraints on the ionization fraction at the tail-end of reionization.
\end{abstract}

\begin{keywords}
galaxies: intergalactic medium--galaxies: high-redshift--cosmology: dark ages, reionization, first stars--galaxies: quasars: absorption lines
\end{keywords}

\section{Introduction}

The detection of quasars at high redshifts $z \sim 6$ enabled a novel way of studying the end stages of reionization of neutral hydrogen (HI) by the early star formation \citep{Fan2000,Fan2001,Fan2002,Fan2003,Fan2004,Songaila2004,Fan2006,Fan2006b}. The Lyman-$\alpha$ (Ly$\alpha$) absorption spectra of these quasars are expected to contain information on the distribution of HI in the intergalactic medium (IGM) along the lines of sight.  These observations of the Ly$\alpha$ optical depth, combined with high-quality numerical simulations, allowed one to estimate the HI photoionization rate $\Gamma_{\mathrm{HI}}$ and thus the number of ionizing photons available in the IGM \citep{Fan2006b,Bolton2007,Calverley2011,Wyithe2011}. More detailed studies of these spectra based on, e.g., the damping wings and near zones \citep{Wyithe2004,Maselli2007,Bolton2007,Wyithe2011,Bolton2011,Greig2017,Eilers2017,Eilers2018b,Durovcikova2020,Davies2020a}, evolution of the IGM temperature \citep{Raskutti2012,Boera2019}, fraction of ``dark'' pixels in the spectra \citep{McGreer2011,McGreer2015}, dark gap statistics \citep{Songaila2002,Gallerani2006,Gallerani2008}, have revealed a wealth of information on reionization.

The constraints obtained on the global HI fraction from these studies were relatively straightforward to interpret and to implement in semi-analytical models \citep{Wyithe2003,Choudhury2005,Pritchard2010,Mitra2011,Mitra2012}. When combined with other observations, e.g., the Thomson scattering optical depth of the Cosmic Microwave Background (CMB) photons \citep{Hinshaw2013,PlanckCollaboration2019}, these models were able to constrain the reionization history to a significant extent. Being analytical or semi-analytical in nature, probing a wide range of parameter space was natural for these models and hence they could be coupled to advanced statistical techniques, e.g., Markov chain Monte Carlo (MCMC) \citep{Mitra2011,Mitra2012,Greig2015,Greig2019}. Overall, the data seemed to be consistent with a picture wherein the reionization was completed by $z \sim 5.8$.

More recently, the Ly$\alpha$ effective optical depth $\tau_{\mathrm{eff}}$ of the quasar absorption spectra at $5.5 < z < 6$, when averaged over large scales ($50 h^{-1}$~cMpc), showed significant fluctuations \citep{Becker2015,Bosman2018,Eilers2018,Eilers2019}. These fluctuations could not be explained by simple models of uniform $\Gamma_{\mathrm{HI}}$ and thus led to various extensions to the existing picture of that time. These included, e.g., temperature fluctuations in the IGM \citep{DAloisio2015}, presence of an undetected population of faint quasars \citep{Chardin2015,Chardin2017}, fluctuations in the mean free path $\lambda_{\mathrm{mfp}}$ of ionizing photons \citep{Davies2016}, presence of HI islands left over from reionization \citep{Nasir2019}, shot noise in the placement of bright sources like quasars \citep{Meiksin2020}. The state of the IGM at these redshifts has been modelled extensively through radiative transfer in the numerical simulations of \citet{Kulkarni2019,Keating2019}, where the essential features of many of the other models have been incorporated in a self-consistent manner (e.g., the mean free path and temperature fluctuations, and the presence of left over neutral islands).

In case the $\tau_{\mathrm{eff}}$ fluctuations are indeed due to the neutral islands, then these fluctuations are directly probing the tail end of reionization. Hence, these observations need to be taken into account while attempting to constrain the reionization history. As mentioned above, \citet{Kulkarni2019} have modelled these neutral patches using high-resolution SPH simulations and a cosmological radiative transfer code. Such simulations are usually computationally expensive and hence are not suited for probing the parameter space. Using semi-analytical or semi-numerical models to constrain reionization has the advantage that one is able to identify all possible histories allowed by the data by varying the free parameters, and subsequently study the state of the IGM along with the properties of the ionizing sources \citep[e.g., cooling, feedback, escape of ionizing photons;][]{Mitra2013,Mitra2018}. In case one wants to include the $\tau_{\mathrm{eff}}$ fluctuation data in such statistical analyses, it becomes imperative to devise ways to model the HI islands in a computationally efficient manner. 

The main aim of this work is to build a model of reionization and Ly$\alpha$ forest at $z \sim 5.5$ which is computationally efficient and hence can be used for parameter space exploration. To achieve this, we use a previously developed semi-numerical method to generate ionized regions, driven by Lyman-continuum photons from galaxies, within relatively low-resolution but large simulation volumes \citep{Choudhury2018}. Once the distribution of the ionized (and neutral) regions is generated, we then model the Ly$\alpha$ optical depth of neutral hydrogen (as would be imprinted on spectra of background point sources such as quasars) using the so-called fluctuating Gunn-Peterson approximation \citep{Croft1998}. The resulting realizations of the quasar absorption spectra are then used for calculating the $\tau_{\mathrm{eff}}$ along large lines of sight to allow proper comparison with observations. Because of the simplifications employed, we need to introduce a few free parameters in the model, thus decreasing its predictive power as compared to radiative transfer simulations. The free parameters are constrained by comparing the model predictions with the observational data, the analysis being possible due to the computational efficiency of our algorithm. The end result of the analysis is that we obtain the range in reionization histories at $5 \lesssim z \lesssim 6$ that are statistically allowed by the data.

The paper is organized as follows: We discuss our method of calculating the Ly$\alpha$ optical depth in \secn{sec:method}. In \secn{sec:results}, we present the main results of our analysis before summarizing and discussing the future outlook in \secn{sec:discussion}. The Appendices are devoted to exploring the model parameters beyond their default values and testing the convergence of our results with respect to the resolution. The cosmological parameters used in this work are $\Omega_m = 0.308, \Omega_{\Lambda} = 1- \Omega_m, \Omega_b = 0.0482, h = 0.678, n_s = 0.961, \sigma_8 = 0.829$ \citep{PlanckCollaboration2014}.

\section{Method}
\label{sec:method}

\subsection{Generation of ionization maps using \texttt{SCRIPT}}
\label{subsec:ionization_maps}

The ionization maps needed for this work are generated using the semi-numerical method introduced in \citet{Choudhury2018}. The method consists of two steps. In the first, we use a collisionless $N$-body simulation to generate the large-scale smoothed density fields, which are then used for generating the large-scale distribution of the collapsed haloes through a sub-grid prescription. In the second step, the density and the halo fields are used as input to an explicitly photon-conserving semi-numerical formalism to generate the distribution of ionized regions.

For the $N$-body simulation, we use the publicly available code GADGET-2\footnote{\texttt{https://wwwmpa.mpa-garching.mpg.de/gadget/}} \citep{Springel2005} and simulate a box of length $256 h^{-1}$~cMpc with $512^3$ particles. The initial conditions for the simulation are generated using the N-GenIC code\footnote{\texttt{https://wwwmpa.mpa-garching.mpg.de/gadget/right.html\#ICcode}}. At redshifts of interest, the simulation outputs in the form of the particle positions are smoothed using a Cloud-in-Cell (CIC) algorithm to generate the matter overdensity field $\Delta_i=\rho_i/\bar\rho$ in a uniform grid with cells labelled by $i$.

Since the particle resolution of our box is not sufficient to identify the collapsed haloes of interest, we employ a sub-grid scheme to compute the large-scale halo distribution from the density field. Given the overdensity field, we use the conditional mass function from ellipsoidal collapse \citep{Sheth2002}, with parameters calibrated to match simulation results, to generate the fraction of mass $f_{\mathrm{coll}, i}$ in collapsed haloes above mass $M_{\mathrm{min}}$ inside every grid cell. Note that this approximate way of computing the collapsed mass works only for relatively larger grid volumes, hence we do not use grids finer than $2 h^{-1}$~cMpc. Our method not only produces the halo mass function consistent with $N$-body simulations \citep{Jenkins2001}, but also the large-scale clustering of haloes.

The generation of ionization maps requires computing two numbers in every cell in the box. The first is the number of hydrogen atoms which is assumed to follow the dark matter at scales of our interest
\be
N_{H, i} = \bar{n}_H~V_{\mathrm{cell}}~\Delta_i,
\ee
where $\bar{n}_H$ is the mean comoving hydrogen number density and $V_{\mathrm{cell}}$ is the comoving volume of the grid cells. Secondly, we need the cumulative number of ionizing photons produced, which can be assumed to be proportional to the mass within collapsed haloes above a mass $M_{\mathrm{min}}$ and is given by
\be
N_{\mathrm{ion}, i} = \zeta~N_{H, i}~f_{\mathrm{coll}, i} = \zeta~\bar{n}_H~V_{\mathrm{cell}}~\Delta_i~f_{\mathrm{coll}, i},
\ee
where $\zeta$ is the ionizing efficiency. 

The ionization maps are generated using the photon-conserving semi-numerical scheme introduced in \citet{Choudhury2018}, which is named \texttt{SCRIPT} (\textbf{S}emi-numerical \textbf{C}ode for \textbf{R}e\textbf{I}onization with \textbf{P}ho\textbf{T}on-conservation). The algorithm consists of two steps: in the first step, we generate ionized ``bubbles'' around individual grid cells (or sources, as the case may be) allowing the cells where multiple bubbles overlap to be ``over-ionized''. In the second step, we deal with the over-ionized cells in the overlapped bubbles by distributing the excess photons to nearby cells. The code is developed primarily for studying the redshifted 21~cm signal from HI during reionization and includes several improvements over existing semi-numerical codes based on the excursion-set formalism, namely, (i) it conserves the number of ionizing photons thus fixing a known shortcoming of earlier models \citep[see, e.g.,][]{Zahn2007,Paranjape2016}, and (ii) consequently ensures the numerical convergence of large-scale properties of the ionization field with respect to the resolution at which the maps are made. The code has recently been optimized so that it takes $2-4$ seconds to complete on a single processor for a $128^3$ grid and $\sim 0.01$ seconds for a $32^3$ grid.

\subsection{Calculation of the photoionization rate within ionized regions}
\label{sec:photoionization}

The output of the semi-numerical method described above provides the neutral hydrogen fraction $x_{\mathrm{HI}}$ for each cell which can, in principle, be zero in regions that are completely ionized. In reality, however, the recombinations will ensure that there exist some residual neutral atoms even in these cells. The fraction of such neutral atoms would be exceedingly small (unless they are in high-density self-shielded regions) and hence would hardly affect the large-scale properties of the 21~cm signal (which was the original motivation for developing \texttt{SCRIPT}). On the other hand, for studying the quasar spectra, this residual neutral hydrogen in the low-density IGM would cause the Ly$\alpha$ absorption observed in quasar spectra and hence needs to modelled properly.

The main ingredient in modelling the residual neutral fraction is the photoionization rate $\Gamma_{\mathrm{HI}}$. To do so,  let us start with the flux incident on the $i$th cell
\be
J_i(\nu) = \sum_{j \neq i} \f{L_j(\nu)}{(4 \pi)^2 (a~x_{ij})^2}~\mathrm{e}^{-\tau_{ij}(\nu)},
\ee
where $L_j(\nu)$ is the luminosity of the $j$th cell, $x_{ij}$ is the \emph{comoving} distance between the $i$th and the $j$th cells and $\tau_{ij}$ is the optical depth between the two cells. The summation extends over all cells other than the cell under consideration. The luminosity $L_i(\nu)$ is related to the quantity $N_{\mathrm{ion}, i}$ used for generating the ionization maps in \secn{subsec:ionization_maps}. However, since $N_{\mathrm{ion}, i}$ is the cumulative number of photons produced, we need to introduce a characteristic time-scale $t_*$ to relate it to the instantaneous luminosity. We can then write
\be
L_i(\nu) = \dot{N}_i(\nu)~h \nu \equiv \f{N_i(\nu)}{t_*}~h \nu,
\ee
where $N_i(\nu)$ is the cumulative number of photons produced per unit frequency range and is related to $N_{\mathrm{ion}, i}$ by
\be
N_{\mathrm{ion}, i} = \int_{\nu_\mathrm{HI}}^{\infty} \de \nu~ N_i(\nu),
\ee
where $\nu_\mathrm{HI}$ is the Lyman-limit frequency. 

The optical depth between the two cells $i$ and $j$ can be calculated by integrating along the sight line joining the cell \citep{Davies2016}
\be
\tau_{ij}(\nu) = \int_{x_i}^{x_j} \f{\de x}{\lambda_{\mathrm{mfp}}(\nu, x)}.
\label{eq:tau_ij}
\ee
As shown in \citet{Davies2016}, computing the mean free path self-consistently requires iterative solutions and can be computationally expensive \citep{Hutter2018}. To start with, for ionized regions, we can make the simplifying assumption that the mean free path takes just one value for the whole box and is determined by the typical distance between the self-shielded regions. Let us denote this mean free path by $\lambda_{\mathrm{ss}}$. This approximation is believed to be adequate in the post-reionization universe. This approximation is probably acceptable also for cells within an ionized region as long as its size is significantly larger than $\lambda_{\mathrm{ss}}$. However, the assumption breaks down at early stages of reionization where the mean free path is pre-dominantly determined by the bubble size. 

With this assumption, we can write a simplified expression for the optical depth
\be
\tau_{ij} = \f{x_{ij}}{\lambda_{\mathrm{ss}}(\nu)}.
\ee
The calculation then follows the usual approach outlined in \citet{Davies2016, Hutter2018} and the photoionization rate can be shown to be given by
\be
\Gamma_{\mathrm{HI}, i} = \f{1}{a^2} \f{\alpha}{\alpha + \beta} ~ \f{\sigma_{\mathrm{HI}}(\nu_{\mathrm{HI}})}{t_*} \sum_{j \neq i} N_{\mathrm{ion}, j}~\f{\mathrm{e}^{-x_{ij} / \lambda_{\mathrm{ss}}}}{4 \pi x_{ij}^2},
\label{eq:Gamma_HI_i}
\ee
where $\alpha$ is the spectral index of the ionizing sources, $\beta$ is the spectral index of the hydrogen ionization cross section, and $\sigma_{\mathrm{HI}}$ is the cross section at $\nu = \nu_\mathrm{HI}$. While deriving the above equation, we have assumed that the mean free path is independent of $\nu$ (which is reasonable because the hydrogen ionization cross section is a steeply declining function of $\nu$).

Note that the summation on the right hand side depends only on $x_{ij}$, hence it can be expressed as a sum over contributions from spherical shells around the $i$th cell. Let us write it as
\be
\Gamma_{\mathrm{HI}, i} = \f{1}{a^2} \f{\alpha}{\alpha + \beta} ~ \f{\sigma_{\mathrm{HI}}(\nu_{\mathrm{HI}})}{t_*} \sum_{J} N^{(i)}_{\mathrm{ion}}(r_J)~\f{\mathrm{e}^{-r_J / \lambda_{\mathrm{ss}}}}{4 \pi r_J^2},
\label{eq:Gamma_HI_i_shells}
\ee
where the summation index $J$ is over all spherical shells. The $J$th shell has a comoving radius $r_J$. The quantity $N^{(i)}_{\mathrm{ion}}(r_J)$ is the number of photons contributed by all the grid cells within the $J$th spherical shell and the superscript ${(i)}$ signifies that the spheres are constructed around the $i$th grid cell. The summation over shells in the above equation can be computed using spherical filters and allows the calculation to be extremely efficient computationally.

A major shortcoming of the above formalism is that it does not account for ``shadows'' arising because of the neutral islands. We expect that the cells close to the boundaries of the ionized regions will not receive contribution from sources in the direction of the neutral regions, hence the photoionization rate in these cells should be less than what is given by \eqn{eq:Gamma_HI_i_shells}. \citet{Nasir2019} implement this effect in a direct way by removing contributions from sources whose lines of sight pass through neutral islands. However, such a method based on ray-tracing turns out to be computationally time-consuming in our simulations. Ideally, we would prefer to retain the summation over spherical shells as this allows the method to be computationally efficient.

\begin{figure}
  \includegraphics[width=0.47\textwidth]{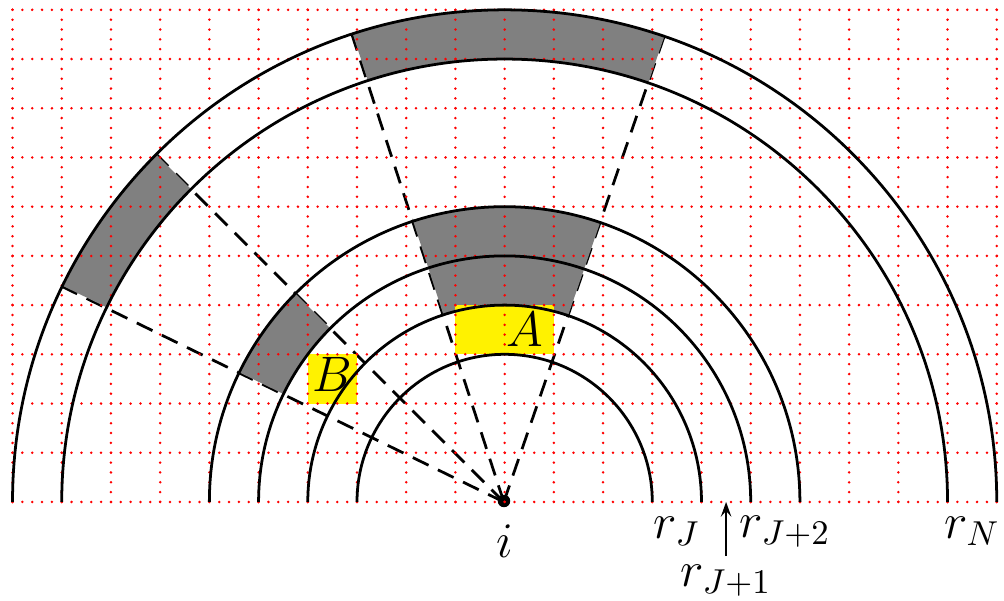}
  \caption{
Illustration of our method of implementing ``shadows'' arising from neutral islands. The simulation box is divided into (cubic) grid cells which are assigned a neutral fraction using semi-numerical method \texttt{SCRIPT}. In this example, the grids in yellow are neutral and all others are assumed to be ionized. The photoionization rate at the grid cell under consideration $i$ is computed by summing over contributions from ionizing sources within spherical shells $J$ having radii $r_J$. The neutral regions will block radiation from sources within the regions marked in gray (showed only for a few representative shells).
  }
  \label{fig:shadows}
\end{figure}

To this end, for every pair of grid cell $i$ and spherical shell $J$, we introduce a correction factor for the photon number $N^{(i)}_{\mathrm{ion}}(r_J)$ to account for the shadows arising from the neutral islands within radius $r_J$ around the $i$th cell. Our method for computing the correction factor is best explained by the illustration in \fig{fig:shadows}. To obtain the photoionization rate $\Gamma_{\mathrm{HI}, i}$, we construct spherical shells of different radii around the $i$th cell and sum over the contribution from each of them as given in \eqn{eq:Gamma_HI_i_shells}. We show a few such shells in the figure. For convenience of implementing the algorithm, we assume all spherical shells to have the same thickness $\Delta r$.

Let us imagine that there are two neutral regions $A$ and $B$ in the volume of interest, the corresponding neutral grid cells being marked in yellow. Although the shapes of the regions look somewhat contrived in the chosen example, our final expression would be written in terms of the fraction of volume occupied by the neutral regions and hence would work even for arbitrary shaped regions.

Clearly, the neutral cells would \emph{not} allow photons to reach $i$ from regions on the other side. For example, the neutral region $A$, which is situated in shell $J$, would block photons from the outer shells $J+1, J+2, \ldots N, \ldots$, the affected regions are marked in gray in the figure. Let the number of neutral cells in shell $J$ be denoted as $N_{\mathrm{cell,HI}}(r_J)$, which in our example would be the number of cells occupied by $A$. The cells in shell $J+1$ that would be affected by these neutral cells would be the ones within the solid angle denoted by the dashed lines joining $i$ and the edges of region $A$. Since all the shells have the same thickness $\Delta r$, the effective number of cells in the $(J+1)$th shell that are affected by the neutral cells in $J$ is approximately given by
\be
N^{(i)}_{\mathrm{cell,shadow}}(r_{J+1}) = N^{(i)}_{\mathrm{cell,HI}}(r_J) ~ \f{r_{J+1}^2}{r_J^2},
\ee
as is obvious from the figure. The above relation is approximate due to edge effects of fitting cubes inside spherical shells. We can now work out, under the same approximation, the effective number of cells affected by the neutral cells in shell $J+2$ 
\bear
N^{(i)}_{\mathrm{cell,shadow}}(r_{J+2}) &= N^{(i)}_{\mathrm{cell,shadow}}(r_{J+1}) ~ \f{r_{J+2}^2}{r_{J+1}^2} 
\nline
&+ N^{(i)}_{\mathrm{cell,HI}}(r_{J+1}) ~ \f{r_{J+2}^2}{r_{J+1}^2},
\ear
where the first term on the right hand side is arising from region $A$ and the second term from $B$. 

It is straightforward to generalize the relation and write down the equivalent formula for any given shell at $r_N$
\bear
N^{(i)}_{\mathrm{cell,shadow}}(r_{N}) &= \left[N^{(i)}_{\mathrm{cell,HI}}(r_{N-1}) \right.
\nline
&\left. + N^{(i)}_{\mathrm{cell,shadow}}(r_{N-1}) \right] \f{r_N^2}{r_{N-1}^2}.
\ear
Since the total number of cells in a spherical shell $N_{\mathrm{cell,tot}}(r_N) \propto r_N^2 \Delta r$, and since $\Delta r$ is chosen to be the same for all shells, the fraction of cells in the $N$th shell that are affected by the neutral islands is given by
\be
Q^{(i)}_{\mathrm{shadow}}(r_{N}) = \mathrm{min}\left[Q^{(i)}_{\mathrm{HI}}(r_{N-1}) + Q^{(i)}_{\mathrm{shadow}}(r_{N-1}), 1\right],
\ee
where $Q^{(i)}_{\mathrm{HI}}(r_{N})$ is the fraction of neutral cells in the $N$th shell. As we consider contributions from shells farther away from the $i$th cell, the fraction $Q^{(i)}_{\mathrm{shadow}}(r_{N})$ keeps on increasing (or remains the same) depending on how the neutral islands are distributed. This allows us to account for the shadows by simply decreasing the photon contribution from the $N$th shell by a factor $N^{(i)}_{\mathrm{ion}}(r_N) \longrightarrow \left[1 - Q^{(i)}_{\,\mathrm{shadow}}(r_N)\right]~N^{(i)}_{\mathrm{ion}}(r_N)$. Since the equation is written solely in terms of neutral volume fractions $Q^{(i)}_{\mathrm{HI}}(r_{N})$, it automatically allows for partially ionized cells (which may arise because the boundaries of the neutral regions can be of arbitrary shapes).

With the above modification, our new relation for the photoionization rate is given by
\bear
\Gamma_{\mathrm{HI}, i} &= \f{1}{a^2} \f{\alpha}{\alpha + \beta} ~ \f{\sigma_{\mathrm{HI}}(\nu_{\mathrm{HI}})}{t_*} \times
\nline
& \times \sum_{J} \left[1 - Q^{(i)}_{\mathrm{shadow}}(r_J) \right]~N^{(i)}_{\mathrm{ion}}(r_J)~\f{\mathrm{e}^{-r_J / \lambda_{\mathrm{ss}}}}{4 \pi r_J^2}.
\label{eq:Gamma_HI_i_shadow}
\ear
Since the algorithm operates at the level of spherical shells (instead of individual cells), it effectively penalizes all sources that happen to lie in a shell. This is clearly only an approximation, since it ignores all direction dependence in the placement of neutral islands and sources around the cell in question. For example, the method tends to double-count the contribution from the neutral regions that lie in the shadow of another neutral region. Nevertheless, we expect the spherical averaging to level some of these discrepancies. Our method still gives lower fluxes at cells that are close to the neutral islands and hence are likely to be affected by shadowing. It also introduces additional fluctuations in the photoionization field within the ionized regions. The main advantage of the method is that it is computationally much faster than any method that depends on lines of sight.

A final point to note is that the summation in \eqn{eq:Gamma_HI_i_shadow} accounts only for the cells other than the cell under consideration. For the local contribution, we assume that the sources within $r_0$ (the radius of the sphere corresponding to the grid volume) are distributed uniformly, hence the photoionization rate is given by \citep{Davies2016}
\be
\Gamma^{\mathrm{local}}_{\mathrm{HI}, i} = \f{1}{a^2} \f{\alpha}{\alpha + \beta} ~\f{\sigma_{\mathrm{HI}}(\nu_{\mathrm{HI}})}{t_*} N_{\mathrm{ion}, i} ~ \left(1 - \mathrm{e}^{- r_0 / \lambda_{\mathrm{ss}}} \right)~\f{3 \lambda_{\mathrm{ss}}}{4 \pi r_0^3}.
\label{eq:Gamma_HI_i_local}
\ee
The final photoionization rate is calculated by adding this local contribution to the one computed using \eqn{eq:Gamma_HI_i_shadow}.

We can write the expression for $\Gamma_{\mathrm{HI}, i}$ in a more useful form as
\be
\f{\Gamma_{\mathrm{HI}, i}}{10^{-12}~\mathrm{s}^{-1}} = A_{\Gamma}~\left(\f{1 + z}{6.5}\right)^2 \f{S_i}{2.37 \times 10^{18}~\mathrm{cm}^{-2}},
\label{eq:Gamma_HI_A_Gamma}
\ee
where
\be
A_{\Gamma} \equiv \left(\f{\alpha}{3}\right)~\left(\f{6}{\alpha + \beta}\right)~\left(\f{10^7~\mathrm{yr}}{t_*}\right)
\ee
is a normalization factor and
\bear
S_i &\equiv \sum_{J} \left[1 - Q^{(i)}_{\mathrm{shadow}}(r_J) \right]~N^{(i)}_{\mathrm{ion}}(r_J)~\f{\mathrm{e}^{-r_J / \lambda_{\mathrm{ss}}}}{4 \pi r_J^2}
\nline
&+ N_{\mathrm{ion}, i} ~ \left(1 - \mathrm{e}^{- r_0 / \lambda_{\mathrm{ss}}} \right)~\f{3 \lambda_{\mathrm{ss}}}{4 \pi r_0^3}
\ear
is the cumulative photon flux at cell $i$. Our method of calculating $\Gamma_{\mathrm{HI}, i}$ thus depends on two parameters, namely, $\lambda_{\mathrm{ss}}$ and $A_{\Gamma}$ (which, in turn, depends on several other parameters $\alpha$, $\beta$ and $t_*$). We shall return to discuss these parameters in \secn{sec:free_params}.

\subsection{The Lyman-$\alpha$ optical depth}

Having calculated the distribution of HI in the IGM, as caused by the Lyman-continuum photons from galaxies, we now compute the Ly$\alpha$ optical depth $\tau_{\alpha}$ arising from the HI field. This $\tau_{\alpha}$ field would get imprinted on the spectra of distant quasars which act as background sources.

For a cell that is identified as completely ionized by our semi-numerical model of reionization, the residual neutral hydrogen fraction is obtained assuming photoionization equilibrium
\be
x_{\mathrm{HI}, i}~\Gamma_{\mathrm{HI}, i} = \f{\chi_\mathrm{He}}{a^3}~\alpha_B(T_i)~n_{H, i}~(1 - x_{\mathrm{HI}, i})^2,
\ee
where $\chi_\mathrm{He} \approx 1.08$ accounts for the excess electron produced by singly-ionized helium, $\alpha_B(T)$ is the case-B recombination rate and the factor $a^3$ accounts for the fact that the number densities used are in comoving units. The solution to the above quadratic equation is straightforward provided we assume a relation between the temperature $T_i$ of the cell and the density, which is usually taken to be a power-law $T_i = T_0~\Delta_i^{\gamma-1}$, where $T_0$ is the temperature of the cell at the mean density and $\gamma$ is the slope. In numerical simulations where reionization is assumed to be instantaneous and uniform, the value of $\gamma$ is found to be around unity right after the reionization is completed and subsequently approaches a value $\sim 1.5$ \citep{Puchwein2015,Gaikwad2018}. In reality, the temperature distribution could be more complicated given that different points in the IGM get reionized at different times and hence the $T_i-\Delta_i$ relation is not necessarily one-to-one. Note that the value of $\gamma$ used in the above applies to pixels that are ionized. If we make a simplifying assumption that a substantial fraction of them got ionized sufficiently early, we can take a one-to-one $T_i-\Delta_i$ relation in such regions. One should also realize that the relation is defined for the $\Delta_i$ field smoothed over some grid size, hence we expect $\gamma$ to depend on the resolution used for carrying out the analysis.

For most cases of interest, applying the above photoionization equilibrium equation to the completely ionized cells yields neutral fractions much smaller than unity which turn out to be
\be
x_{\mathrm{HI}, i} \approx \f{\chi_\mathrm{He}}{a^3}~\f{\alpha_B(T_i)~n_{H, i}}{\Gamma_{\mathrm{HI}, i}}.
\ee
The above relation is applied only to those cells that are identified as completely ionized by the semi-numerical method, whereas for cells that are partially or completely neutral, we assign the neutral fraction as obtained from the semi-numerical calculation itself.

Under the fluctuating Gunn-Peterson approximation, the Ly$\alpha$ optical depth is given by 
\be
\tau_{\alpha, i} = \kappa_{\mathrm{res}} \f{\pi \e^2}{m_e c} f_{\alpha}~\lambda_{\alpha}~H^{-1}(z)~x_{\mathrm{HI}, i} \f{n_{H, i}}{a^3},
\ee
where $\kappa_{\mathrm{res}}$ is a normalization factor to account for the small-scale fluctuations in the density and velocity fields that are not resolved in our coarse-resolution simulations \citep{Dixon2009,Davies2016}, $f_{\alpha}$ is the Ly$\alpha$ oscillator strength and all other symbols have their usual meanings. A straightforward calculation shows that the optical depth can be written as
\bear
\tau_{\alpha, i} &= 5.01~\kappa_{\mathrm{res}}~\left(\f{\chi_\mathrm{He}}{1.08}\right)~\left(\f{1 - Y}{0.76}\right)^2~\left(\f{\Omega_b h^{3/2}}{0.0269}\right)^2
\nline
&\times \left(\f{1 + z}{6.5}\right)^6~\left(\f{9.23}{H(z)/H_0}\right)
\nline 
&\times \left(\f{T_0}{10^4~\mathrm{K}}\right)^{-0.7}
~\left(\f{10^{-12}~\mathrm{s}^{-1}}{\Gamma_{\mathrm{HI}, i}}\right) ~ \Delta_{i}^{2.7 - 0.7 \gamma},
\label{eq:tau_alpha_Gamma_HI}
\ear
where we have assumed $\alpha_B(T) \propto T^{-0.7}$. A more useful form can be obtained by substituting $\Gamma_{\mathrm{HI}, i}$ from \eqn{eq:Gamma_HI_A_Gamma}
\bear
\tau_{\alpha, i} &= 5.01~A_{\tau}~\left(\f{\chi_\mathrm{He}}{1.08}\right)~\left(\f{1 - Y}{0.76}\right)^2~\left(\f{\Omega_b h^{3/2}}{0.0269}\right)^2
\nline
&\times \left(\f{1 + z}{6.5}\right)^4~\left(\f{9.23}{H(z)/H_0}\right)
\nline
&\times \f{2.37 \times 10^{18}~\mathrm{cm}^{-2}}{S_i}~ \Delta_{i}^{2.7 - 0.7 \gamma},
\label{eq:tau_alpha}
\ear
where we define a new normalization factor
\bear
A_{\tau} &\equiv \kappa_{\mathrm{res}}~\left(\f{T_0}{10^4~\mathrm{K}}\right)^{-0.7}~A_{\Gamma}^{-1}
\nline
&= \kappa_{\mathrm{res}}~\left(\f{T_0}{10^4~\mathrm{K}}\right)^{-0.7}
~\left(\f{3}{\alpha}\right)~\left(\f{\alpha + \beta}{6}\right)~\left(\f{t_*}{10^7~\mathrm{yr}}\right).
\label{eq:A_tau}
\ear

The transmitted flux for the cell is given by ${\rm e}^{-\tau_{\alpha, i}}$, hence the effective optical depth averaged over $N$ pixels is given by
\be
\tau_{\mathrm{eff}} = -\ln \left[\f{1}{N} \sum_{i} {\rm e}^{-\tau_{\alpha, i}}  \right].
\ee
This is the main observable in our work which will be compared with the observations. 

It is obvious from \eqn{eq:tau_alpha} that it is \emph{not} possible to constrain the physical quantities $\alpha$, $\beta$, $t_*$ and $T_0$ individually from observations of the optical depth, we can only hope to constrain the combination $A_{\tau}$. Additionally to be noted is that the value of $A_{\Gamma}$ is completely degenerate with $T_0$ and $\kappa_{\mathrm{res}}$, therefore we cannot measure the amplitude of $\Gamma_{\mathrm{HI}}$ from our low-resolution simulations. The fluctuations in the rate, however, should be correctly captured in our model.

\subsection{Model parameters}
\label{sec:free_params}

Our model has \emph{five} parameters (which, in general, can be functions of $z$):
\begin{enumerate}

\item The first parameter is the ionizing efficiency $\zeta$ which is used for generating the ionization maps. We treat this as a free parameter.

\item The other parameter that is required for generating the ionization maps is the minimum mass $M_{\mathrm{min}}$ of haloes that are capable of producing ionizing photons. In this work, we choose $M_{\mathrm{min}} = 10^9~\Msun$ which is appropriate for late stages of reionization. At these redshifts, most regions are photoheated and hence the star-formation threshold is set by the radiative feedback. We have also varied $M_{\mathrm{min}}$ in the range $10^8 - 10^{10}~\Msun$ and found that our constraints on the reionization history are insensitive to the value of $M_{\mathrm{min}}$, see Appendix \ref{app:Mmin}.

\item The next parameter is the mean free path $\lambda_{\mathrm{ss}}$ of ionizing photons as determined by the distance between the self-shielded regions. We choose its value as extrapolated from $z \lesssim 5$ observations \citep{Worseck2014} having the empirical power-law fitting form 
\be
\lambda_{\mathrm{ss}}(z) = 175~\mathrm{cMpc}~ \left(\f{1 + z}{5.0} \right)^{-4.4}.
\label{eq:lambda_ss}
\ee
We study the effect of $\lambda_{\mathrm{ss}}$ values different from the above default choice in Appendix \ref{app:mfp}.

\item The fourth parameter in our list is the slope $\gamma$ of the temperature-density relation in the IGM. We leave it as a free parameter.

\item The final parameter is the normalization factor $A_{\tau}$ used for computing the Ly$\alpha$ optical depth, see \eqn{eq:A_tau}. This too is kept free. Since we would want to vary this parameter over orders of magnitude, we prefer $\log A_{\tau}$ as the free parameter while carrying out the statistical analysis.

\end{enumerate}

To summarize, we treat $\zeta$, $\gamma$ and $\log A_{\tau}$ as free parameters and constrain them by comparing with Ly$\alpha$ optical depth data. We fix the values of $M_{\mathrm{min}}$ and $\lambda_{\mathrm{ss}}$ to default values (and check the effect of varying them in Appendix \ref{app:parameters}). We also emphasize here that the constraints on $\gamma$ and $A_{\tau}$ are expected to depend on the resolution used for the analysis, hence it is important to check whether our conclusions regarding the reionization history remain unchanged with respect to the resolution. We investigate this aspect in Appendix \ref{app:resolution}.

\subsection{Observational data}

The main observational data used in this work is from \citet{Bosman2018} who have measured the Ly$\alpha$ effective optical depth $\tau_{\mathrm{eff}}$ averaged over $50 h^{-1}$~cMpc chunks in the redshift range $5 \lesssim z \lesssim 6$. We use their `GOLD' sample which consists of spectra with various quality-cuts. Their results are presented as `optimistic' and `pessimistic' limits on the cumulative distribution function (CDF) $P(<\tau_{\mathrm{eff}})$ depending on how they treat the non-detections of the transmitted flux. In the optimistic case, the lower limits on $\tau_{\mathrm{eff}}$ are treated as measurements just below the detection sensitivity, while in the pessimistic case, these are assumed to have $\tau_{\mathrm{eff}} \to \infty$.

\section{Results}
\label{sec:results}

\subsection{Constraints on the reionization history}
\label{sec:constraints}

We now compare the predictions of our model with the observational data on $\tau_{\mathrm{eff}}$ fluctuations to constrain the reionization history. We treat each redshift bin as independent and constrain the three free parameters of our model using a Bayesian likelihood method. The main steps followed in the analysis can be summarized as follows:

\begin{itemize}

\item We first convert the observational data on the CDF $P(<\tau_{\mathrm{eff}})$ to the differential PDF $p(\tau_{\mathrm{eff}}) \equiv \de P(<\tau_{\mathrm{eff}}) / \de \tau_{\mathrm{eff}}$. Using the differential distribution for our analysis ensures that each measurement of $\tau_{\mathrm{eff}}$ from the observed spectra contributes to only one bin thus reducing correlations across different bins.

\item The pessimistic and optimistic limits of \citet{Bosman2018} data differ in the way the non-detections of the transmitted flux are treated. In principle, one can take the forward modelling approach and use the noise characteristics of the telescopes to contaminate the simulated spectra appropriately. This will allow a fair comparison with the data without making any assumptions about the value of $\tau_{\mathrm{eff}}$ in case of non-detections. However, the features most affected while adding the noise are narrow transmission spikes \citep[$\lesssim$~cMpc across, see][]{Chardin18,Gaikwad2020} which are not resolved by the low-resolution pixels of our model. Hence, we take a different approach where we treat the optimistic and pessimistic bounds as two independent data sets and compare them with the simulated spectra without adding any noise. Since the bounds provide reliable extrema for the recovery of the underlying distribution, the two sets of constraints thus obtained on the reionization history should bracket the full range of allowed histories.

\item The likelihood analysis requires computing the $\chi^2$ defined as
\be
\chi^2 = \sum_{\alpha, \beta = 1}^{N_{\mathrm{bins}}} \Delta p(\tau_{\mathrm{eff}, \alpha})~\left[C^{-1}\right]_{\alpha \beta}~\Delta p(\tau_{\mathrm{eff}, \beta}),
\label{eq:chisq}
\ee
where $\tau_{\mathrm{eff}, \alpha}$ is the value corresponding to $\alpha$-bin, $C^{-1}$ is the inverse of the error covariance matrix and 
\be
\Delta p(\tau_{\mathrm{eff}, \alpha}) \equiv p_{\mathrm{model}}\left(\tau_{\mathrm{eff}, \alpha}; \boldsymbol{\theta}\right) - p(\tau_{\mathrm{eff}, \alpha}).
\ee
In the above definition, $p(\tau_{\mathrm{eff}, \alpha})$ is the binned PDF computed from the observational data and $p_{\mathrm{model}}\left(\tau_{\mathrm{eff}, \alpha}; \boldsymbol{\theta}\right)$ is the theoretical PDF for the parameter set $\boldsymbol{\theta} \equiv \left\{\zeta, \gamma, \log A_{\tau}\right\}$. For a given redshift, we compute $p_{\mathrm{model}}\left(\tau_{\mathrm{eff}, \alpha}; \boldsymbol{\theta}\right)$ from our simulations by first drawing as many random lines of sight of length $50 h^{-1}$~cMpc as there are in the observational data and then computing the mean over 5000 independent realizations of the sight lines.

\item We assume that, for a given data set, the errors on the distribution are dominated by the variations across different sight lines. We thus estimate the covariance matrix elements $C_{\alpha \beta}$ from 5000 independent realizations of the PDF from the simulation, accounting for any possible correlations between different $\tau_{\mathrm{eff}}$ bins. The number of realizations is chosen to ensure that the $\chi^2$ is numerically converged. Clearly, the elements $C_{\alpha \beta}$ depend on the parameter values $\boldsymbol{\theta}$ and ideally one should compute them for every point $\boldsymbol{\theta}$ in the parameter space during the Bayesian statistical analysis. This, however, substantially increases the time taken for exploring the parameter space, hence, we compute the covariance matrix only for a fiducial parameter set $\boldsymbol{\tilde{\theta}}$ (for each redshift and each data set) and use it throughout the analysis. The fiducial parameter values $\boldsymbol{\tilde{\theta}}$ are obtained by minimizing the $\chi^2$ in \eqn{eq:chisq}. We use the BOBYQA bounded minimization routine of \citet{Powell2009} for this purpose. During the minimization, we compute $C_{\alpha \beta}$ individually for every point $\boldsymbol{\theta}$ in the parameter space, which is manageable only because the minimization requires substantially less number of evaluations of the $\chi^2$ compared to the full Bayesian analysis.

\item We use the publicly available affine-invariant ensemble sampler for Markov chain Monte Carlo (MCMC) called \texttt{emcee} \citep{Foreman-Mackey2013} to obtain the posterior distribution of the parameters. The sampler requires evaluation of a likelihood for every $\boldsymbol{\theta}$, which we define as ${\cal L} = \e^{-\chi^2 / 2}$. We use flat priors on all the parameters in the ranges given below:
\begin{itemize}
\item[$\star$] $\zeta$ is assumed to have a flat prior in the range $[0, \zeta_{\mathrm{max}}]$, where $\zeta_{\mathrm{max}}$ is the value corresponding to complete ionization of the IGM.
\item[$\star$] $\gamma$ is assumed to have a flat prior in the range $[0, 3]$. This prior is quite conservative and allows for a wide range in thermal states of the IGM (including ``inverted'' temperature-density relations). 
\item[$\star$] $\log A_{\tau}$ is assumed to have a flat prior in the range $[-3, 3]$. 
\end{itemize}
We use 20 walkers and run the chains long enough so that they converge, which is assessed through the auto-correlation analysis of \citet{Goodman2010}.

\item Our default runs are carried out at grid resolution $\Delta x = 8 h^{-1}$~cMpc. Given our box, this leads to $32^3$ grid cells, thus making the code extremely fast. The sensitivity of our results to the grid size is investigated in Appendix~\ref{app:resolution}.

\end{itemize}

\begin{figure}
  \includegraphics[width=0.49\textwidth]{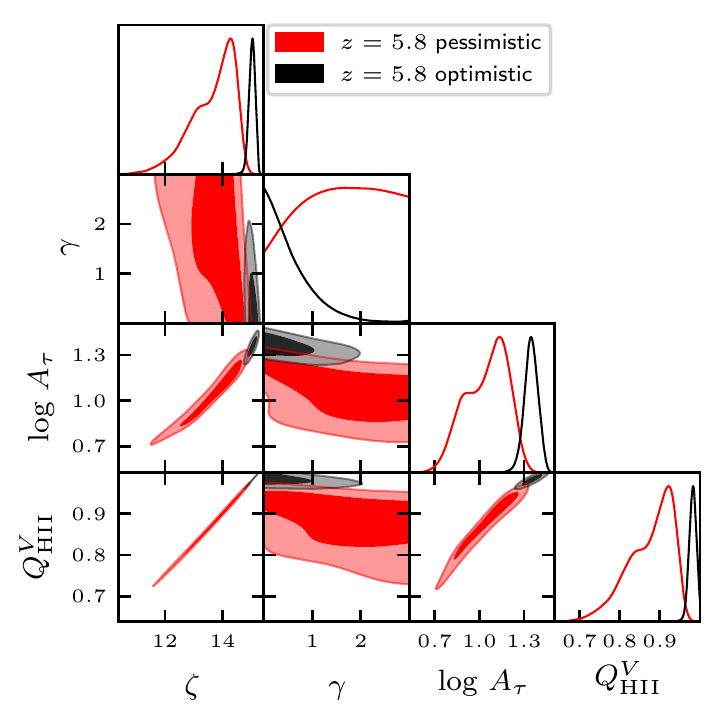}
  \caption{
The marginalized posterior distributions for the free parameters $\zeta$, $\gamma$ and $\log A_{\tau}$ and one derived parameter $Q^V_{\mathrm{HII}}$ obtained from the MCMC analysis. The results are shown at $z = 5.8$ for the pessimistic (red) and optimistic (black) data sets. The contours enclose $68\%$ and $95\%$ of the points.
}
  \label{fig:getdist_onez_triangle}
\end{figure}

\begin{figure}
  \includegraphics[width=0.47\textwidth]{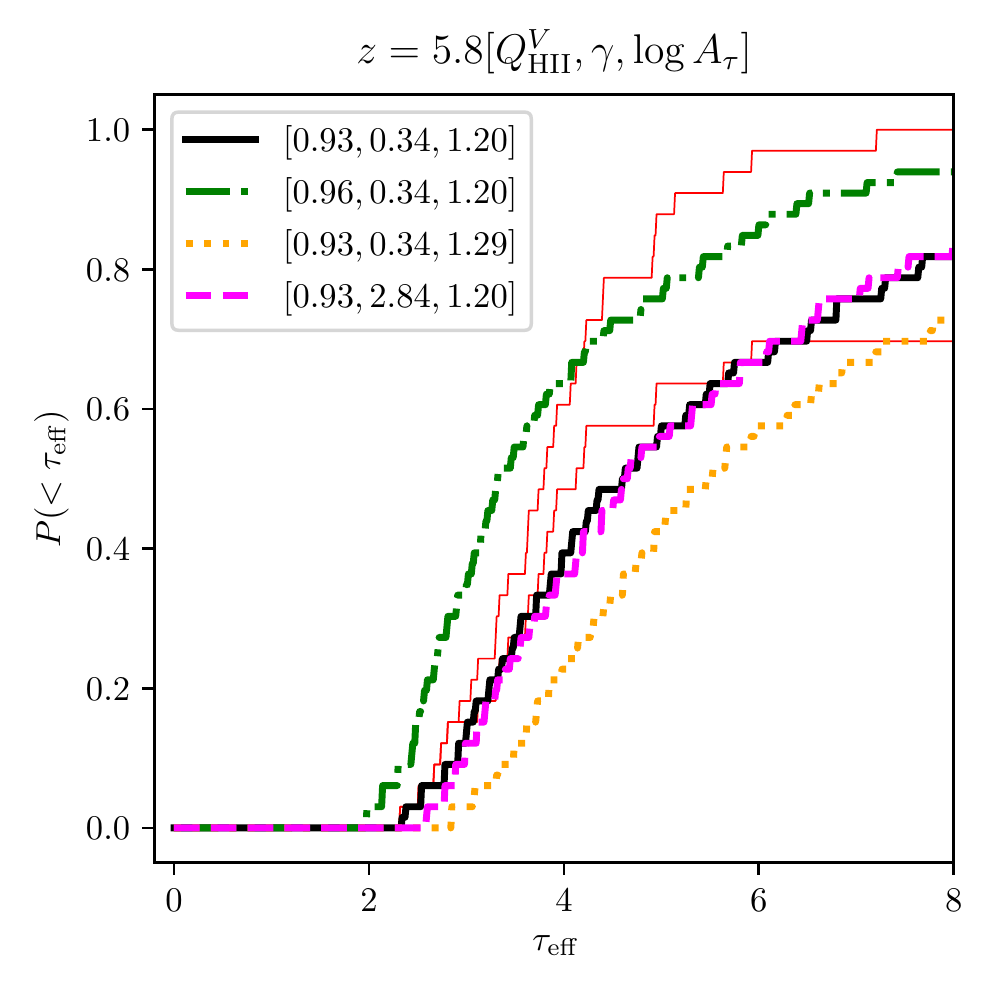}
  \caption{Dependence of the cumulative $\tau_{\mathrm{eff}}$ distribution on model parameters. The two thin red curves show the observational data for the optimistic (left/upper) and pessimistic (right/lower) cases of \citet{Bosman2018}. The black solid curve is the ``default'' model for the purpose of this plot which is essentially the best-fit to the pessimistic data at $z = 5.8$. The green dot-dashed, orange dotted and magenta dashed curves show the variation of the $\tau_{\mathrm{eff}}$ distribution with respect of $Q^V_{\mathrm{HII}}$, $\log A_{\tau}$ and $\gamma$, respectively.}
  \label{fig:plot_taueff_3p_onez}
\end{figure}

The result of our analysis, for one representative redshift $z = 5.8$, is shown in \fig{fig:getdist_onez_triangle} where we plot the marginalized posterior distributions for the three free parameters along with a derived parameter $Q^V_{\mathrm{HII}}$. The distributions are obtained by comparing the model with the optimistic (black) and pessimistic (red) data sets. We can see that the allowed values of $\zeta$, and hence of $Q^V_{\mathrm{HII}}$, are higher for the optimistic data set than the pessimistic one. This is along the expected lines as the inferred values of $\tau_{\mathrm{eff}}$ for sight lines with no detections are higher in the pessimistic case, hence matching this data set requires more neutral patches in the IGM. The statistical errors on the parameters are typically larger in the pessimistic case. This is because the best-fit values of $Q^V_{\mathrm{HII}}$ for the pessimistic case are usually smaller and the presence of more neutral islands introduces larger variations across lines of sight. This naturally leads to larger errors on the $\tau_{\mathrm{eff}}$ distribution (i.e., larger values of the covariance matrix elements $C_{\alpha \beta}$).

\begin{table*}
\begin{tabular}{|c|c|c|c|c|c|c|}
\hline
Parameters: & \multicolumn{2}{c}{$\gamma$} & \multicolumn{2}{c}{$\log A_{\tau}$} & \multicolumn{2}{c}{$Q^V_{\mathrm{HII}}$} \\
\hline
Data type:  & optimistic & pessimistic & optimistic & pessimistic & optimistic & pessimistic \\
\hline
$z = 5.0$ & 0.80 [0.18, 1.82] & 0.80 [0.18, 1.82] & 1.49 [1.31, 1.66] & 1.49 [1.31, 1.66] & 0.999 [$ > 0.996$]   & 0.999 [$ > 0.996$]\\
$z = 5.2$ & 0.94 [0.09, 1.68] & 0.89 [0.15, 2.13] & 1.44 [1.39, 1.60] & 1.44 [1.36, 1.47] & 0.996 [$ > 0.985$]   & 0.997 [$ > 0.983$]\\
$z = 5.4$ & 0.52 [$ < 1.67$]  & 0.69 [$ < 2.55$]  & 1.42 [1.36, 1.45] & 1.43 [1.33, 1.44] & 0.998 [$ > 0.990$]   & 1.000 [$ > 0.984$]\\
$z = 5.6$ & 0.25 [$ < 1.62$]  & 0.45              & 1.37 [1.30, 1.43] & 1.30 [1.24, 1.37] & 0.983 [0.972, 0.993] & 0.969 [0.952, 0.982]\\
$z = 5.8$ & 0.14 [$ < 1.61$]  & 0.34              & 1.35 [1.26, 1.44] & 1.20 [0.78, 1.29] & 0.974 [0.959, 0.993] & 0.926 [0.757, 0.958]\\
$z = 6.0$ & 0.08              & 0.70              & 1.33 [1.31, 1.38] & 0.67 [0.36, 1.00] & 0.982 [0.971, 0.994] & 0.718 [0.561, 0.820]\\
\hline
\end{tabular}
\caption{Best-fit values of the model parameters along with the $2 \sigma$ confidence limits (in parentheses) obtained from the likelihood analysis. Cases for which the $2 \sigma$ range is not mentioned imply that there were no constraints within the prior range considered. $\gamma$ and $\log A_{\tau}$ are the free parameters of the model, while $Q^V_{\mathrm{HII}}$ is a derived parameter that is perfectly correlated with the third free parameter $\zeta$.}
\label{tab:best-fit}
\end{table*}

To further understand the constraints on different parameters, we show in \fig{fig:plot_taueff_3p_onez} the dependence of the observable $P(<\tau_{\mathrm{eff}})$ on the three parameters $Q^V_{\mathrm{HII}}$, $\log A_{\tau}$ and $\gamma$.\footnote{The dependence of $P(<\tau_{\mathrm{eff}})$ on $\zeta$ is very similar to that on $Q^V_{\mathrm{HII}}$ as these two parameters are perfectly correlated, hence we do not show the effect of $\zeta$ separately.} It is clear from the figure that increasing $Q^V_{\mathrm{HII}}$ leads to less number of high opacity sight lines (compare the black solid curve with the green dot-dashed), thus decreasing the scatter in the distribution. This is expected since it is the presence of neutral islands that leads to the high opacity regions. The effect of increasing the normalization $A_{\tau}$ is to increase the value of $\tau_{\mathrm{eff}}$, hence the whole distribution shifts to the right (compare the black solid curve with the orange dotted). Interestingly, the effect of $\gamma$ on the $\tau_{\mathrm{eff}}$ distribution is quite minimal. One can see that in spite of increasing $\gamma$ by $\sim 2.5$, the change in $P(<\tau_{\mathrm{eff}})$ is negligible (compare the black solid curve with the magenta dashed). A closer look at the curves reveals that a higher $\gamma$ tends to increase the value of $\tau_{\mathrm{eff}}$ along lines of sight of relatively lower opacities and vice versa, and hence the distribution becomes slightly narrower. If we ignore the neutral islands for the moment, this can be understood as follows: high density regions tend to remain more neutral because they recombine more efficiently. Increasing $\gamma$, on the other hand, has the opposite effect where it leads to a higher temperature in the high-density regions, thus making them more ionized. So the main effect of increasing $\gamma$ is to reduce the scatter in the neutral hydrogen distribution and hence in the optical depth distribution. This is essentially the reason a higher $\gamma$ leads to a narrower $\tau_{\mathrm{eff}}$ distribution. 

\begin{figure}
  \includegraphics[width=0.47\textwidth]{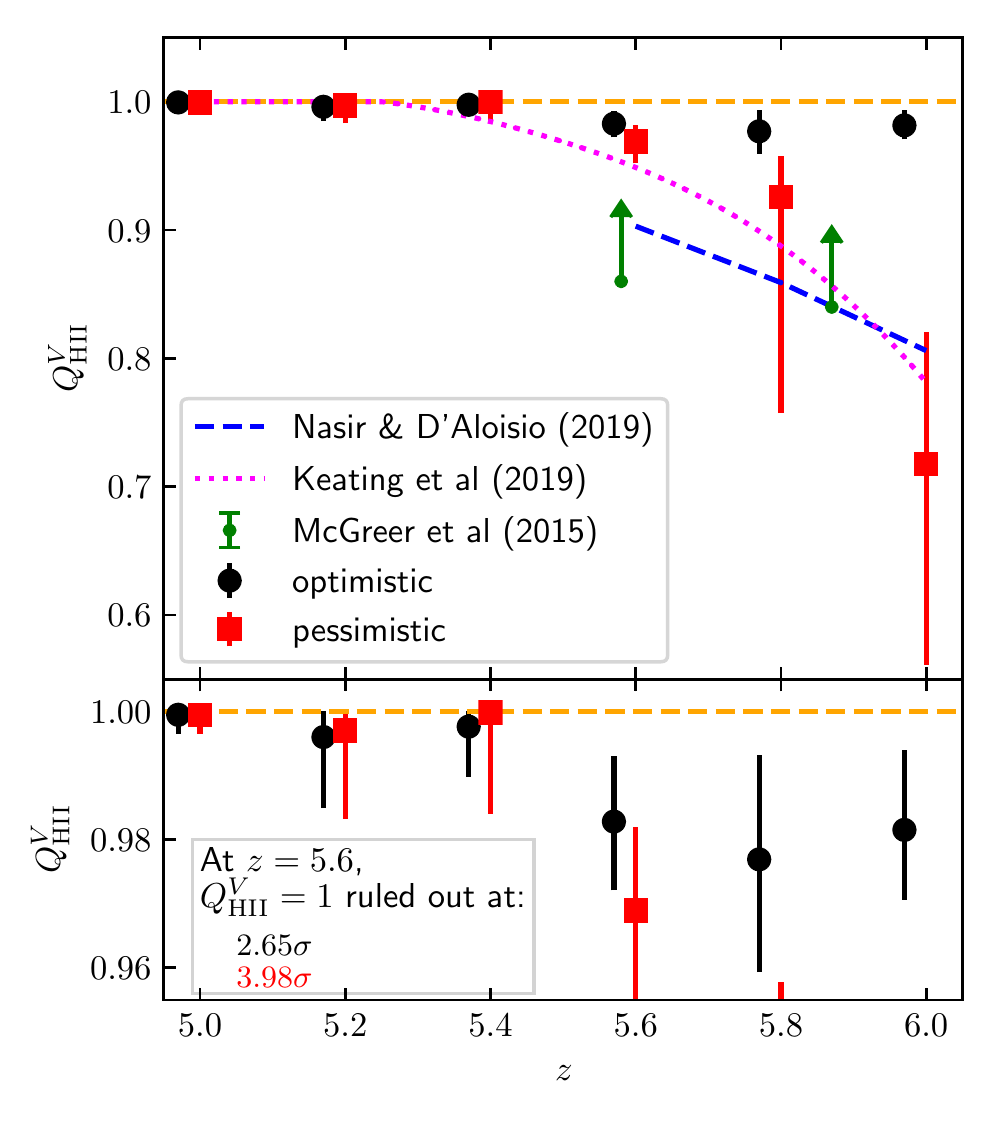}
  \caption{\textbf{Constraints on the ionized volume fraction:} The top panel shows the constraints on $Q^V_{\mathrm{HII}}$ obtained from the MCMC analysis for the optimistic (black circles with error-bars, shifted slightly along the redshift axis for clarity) and pessimistic (red squares with error-bars) data sets of \citet{Bosman2018}. The horizontal orange dashed line corresponds to $Q^V_{\mathrm{HII}} = 1$. The blue dashed curve corresponds to the \texttt{late-reion-long-mfp} reionization history of \citet{Nasir2019} while the magenta dotted curve is the the default reionization model of \citet{Keating2019}. The green points with error-bars reflect the $2\sigma$ lower limits on $Q^V_{\mathrm{HII}}$ obtained using the dark pixel fraction \citep{McGreer2015}. 
  The bottom panel shows our constraints on $Q^V_{\mathrm{HII}}$ (same as in the top panel), but the vertical axis scaled such that the behaviour around $Q^V_{\mathrm{HII}} \sim 1$ can be clearly visualized. It is clear that $Q^V_{\mathrm{HII}} = 1$ is ruled out at $z \geq 5.6$ with a significance $> 2 \sigma$ for either of the data sets.
  }
  \label{fig:Qrange_MCMC}
\end{figure}

Returning to the posterior distribution in \fig{fig:getdist_onez_triangle}, we find that the constraints on $\gamma$ are extremely weak which follows from the fact that $\tau_{\mathrm{eff}}$ distributions are relatively insensitive to $\gamma$ in the probed range. In fact, there is no constraint for the pessimistic case within the prior range chosen. For the optimistic case, the statistical errors on the $\tau_{\mathrm{eff}}$ distribution are smaller and hence the data is able to put some constraints on $\gamma$. In this case, we  can rule out $\gamma > 1.6$ (at $2 \sigma$). This is because for higher values of $\gamma$ the distribution becomes too narrow to be allowed by the data. In general, we found that fixing $\gamma = 1.5$ and carrying out a two-parameter MCMC analysis leave the posteriors of the other parameters unaffected. There is a strong positive correlation between $\zeta$ and $\log A_{\tau}$ and, as a consequence, between $Q^V_{\mathrm{HII}}$ and $A_{\tau}$. Higher $\zeta$-values correspond to more ionized IGM which lead to less neutral islands and hence lower $\tau_{\alpha}$. This can be compensated by increasing $A_{\tau}$ appropriately which thus gives rise to the positive correlation. Note that the main goal of our work is to constrain the reionization history which is achieved essentially from the marginalized posterior of the derived parameter $Q^V_{\mathrm{HII}}$. We find that the correlations between different parameters and other properties of the posteriors are similar for other redshifts, hence we do not show the detailed contour plots separately. Instead, we provide the best-fit values of the parameters $\gamma$, $\log A_{\tau}$ and $Q^V_{\mathrm{HII}}$ along with their $2 \sigma$ allowed range in Table \ref{tab:best-fit}.

The best-fit values of $Q^V_{\mathrm{HII}}$ along with the $2\sigma$ errors obtained from our likelihood analysis for the pessimistic (red squares with error-bars) and optimistic (black circles with error-bars, shifted slightly along the horizontal axis for clarity) cases are shown in \fig{fig:Qrange_MCMC}. The horizontal orange dashed line corresponds to $Q^V_{\mathrm{HII}} = 1$. For reference, we also show the \texttt{late-reion-long-mfp} model of \citet{Nasir2019} (blue dashed curve) and the default reionization model of \citet{Keating2019} (magenta dotted curve, almost identical to the very late model of \citealt{Kulkarni2019} in the redshift range of our interest). The $2\sigma$ lower limits on $Q^V_{\mathrm{HII}}$ obtained using the dark pixel fraction \citep{McGreer2015} are shown by green points with error-bars. Our $Q^V_{\mathrm{HII}}$ limits for the optimistic data set are higher than the simulations of \citet{Keating2019,Nasir2019}. The constraints for the pessimistic data set, on the other hand, are consistent with the reionization history of \citet{Keating2019} within the $2\sigma$ error-bars (except that our $2\sigma$ lower limit is marginally higher than the \citealt{Keating2019} value). However, at $z = 5.6$, the allowed values of $Q^V_{\mathrm{HII}}$ from our analysis are significantly higher than that of \citet{Nasir2019}. This indicates that our method, for the same value of neutral fraction, produces more $\tau_{\mathrm{eff}}$ fluctuations than the model of \citet{Nasir2019}. All our constraints are consistent with the lower limits on  $Q^V_{\mathrm{HII}}$ from the model-independent dark pixel fraction \citep{McGreer2015}.

As expected, the allowed values of $Q^V_{\mathrm{HII}}$ are higher for the optimistic data set than the pessimistic one at $z \geq 5.2$. The differences between the two data sets decrease at smaller $z$ because of fewer non-detections. At $z = 5.6$, the the $2\sigma$ upper limit on $Q^V_{\mathrm{HII}}$ is $0.99$ for the optimistic data set, while it is $0.98$ for the pessimistic one.  In fact, we can rule out $Q^V_{\mathrm{HII}} = 1$ at $3.98\sigma$ ($99.993\%$ confidence) for the pessimistic data and at $2.65\sigma$ ($99.2\%$ confidence) for the optimistic data. This indicates that, in order to match the data, the completion of reionization must be delayed until $z \sim 5.6$ (independent of how the non-detections are treated), i.e., the data are \emph{not} consistent with complete reionization at $z > 5.6$. Interestingly, the $2\sigma$ lower limit on $Q^V_{\mathrm{HII}}$ at $z = 5.2$ is $\approx 0.98$ for both the data sets, which would imply a rather significantly late completion of reionization. We find that our analysis requires the IGM to be completely ionized at $z = 5$ and is able to limit the range of $Q^V_{\mathrm{HII}}$ to within a tight limit of $\sim 0.02$ in the redshift range $5.2 \leq z \leq 5.4$ (irrespective of which data set is used). The constraints at $z = 5.8$ and $z = 6$ for the pessimistic data set are rather weak because of the larger errors on the $\tau_{\mathrm{eff}}$ distribution. One can see that $Q^V_{\mathrm{HII}}$ can be as small as $0.76$ at $z = 5.8$ and $0.56$ at $z = 6$ if the pessimistic data set represents reality, although the lower limits may be underestimated by $\sim 0.06$ because of resolution effects (see Appendix \ref{app:resolution}).

This statistical analysis shows the main benefit of our model. Given its computational efficiency, we can probe the parameter space in a reasonable amount of time and hence determine the range of histories allowed by the data. It is important to keep in mind here that, although our constraints on $Q^V_{\mathrm{HII}}$ are reasonably robust, our model cannot constrain other interesting physical quantities like the temperature and the amplitude of the photoionization rate. With improved data sets in the future, we can expect to test some of the model assumptions more critically and put more stringent constraints on reionization.

\subsection{Aspects of the best-fit reionization history}
\label{sec:best_fit}

\begin{figure*}
  \includegraphics[width=0.9\textwidth]{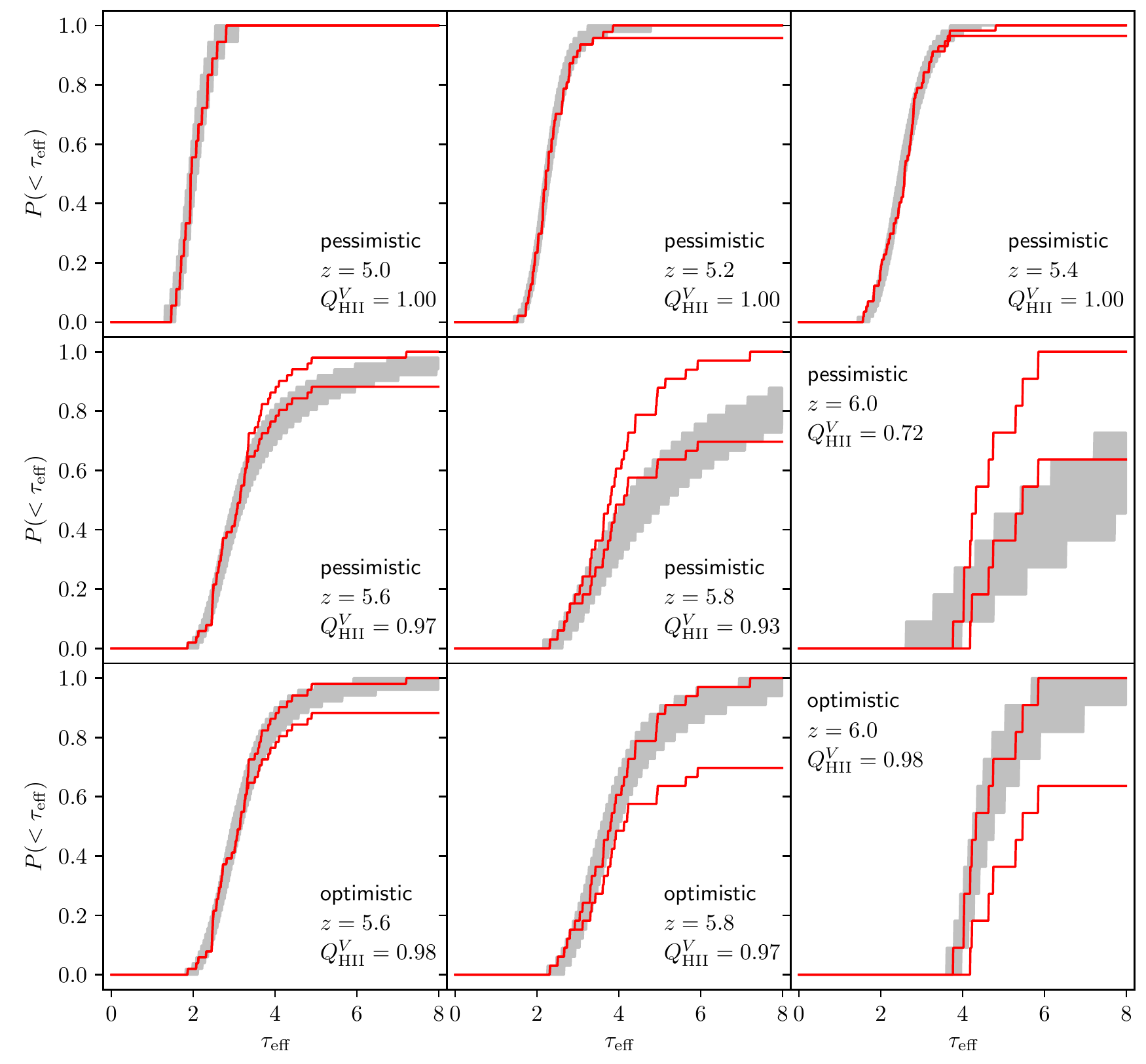}
  \caption{The cumulative $\tau_{\mathrm{eff}}$ distribution for the best-fit reionization model obtained from the MCMC analysis in \secn{sec:constraints}. The two red curves show the observational data for the optimistic (left/upper) and pessimistic (right/lower) cases of \citet{Bosman2018}, with the legend mentioning which of the two data sets is used for model comparison. The gray shaded regions correspond to the model predictions accounting for statistical fluctuations along different sight lines. The redshifts and the ionized volume fractions $Q^V_{\mathrm{HII}}$ are mentioned in the respective panels. Note that we do \emph{not} show the results for the optimistic data set at $z = 5.0$, $5.2$ and $5.4$ as the best-fit models in these cases are almost identical to those for the pessimistic data.}
  \label{fig:taueff}
\end{figure*}

\begin{figure*}
  \includegraphics[width=\textwidth]{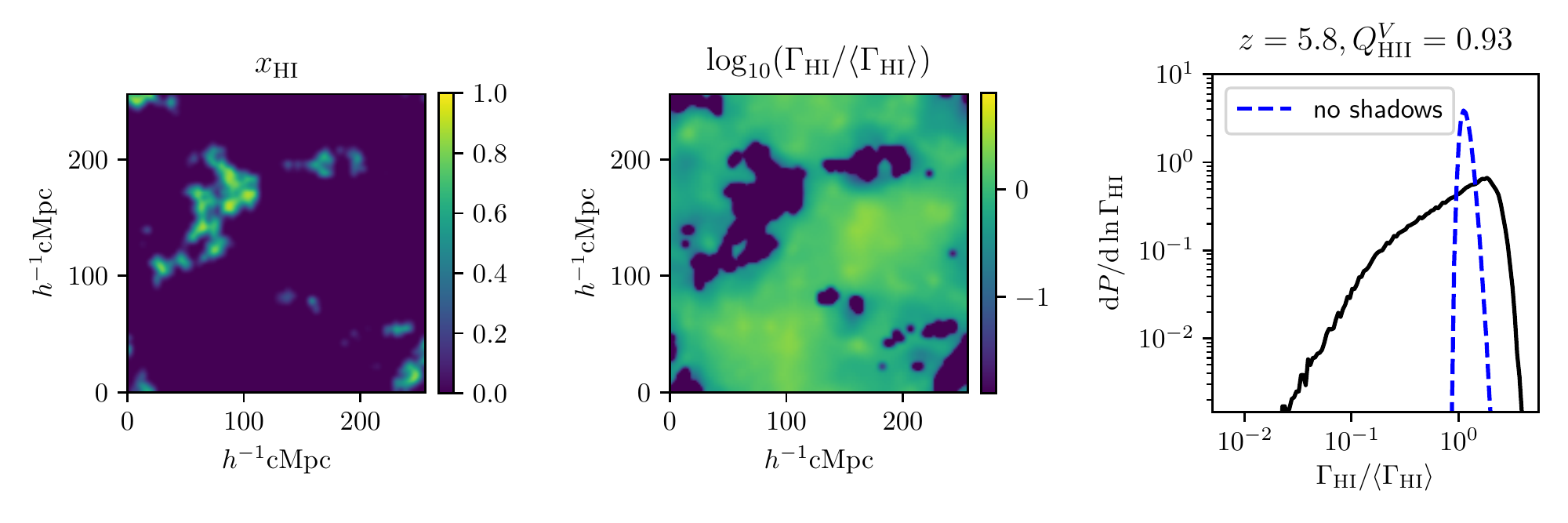}
  \caption{Various quantities obtained from our semi-numerical simulation at $z = 5.8$ where the ionized volume fraction $Q^V_\mathrm{HII} = 0.93$. The left hand panel shows the neutral hydrogen fraction $x_{\mathrm{HI}}$ field for a slice of thickness $4 h^{-1}$~cMpc, while the middle panel shows the $\Gamma_{\mathrm{HI}}$ fluctuations for the same slice. The black curve in the right hand panel shows the PDF of $\Gamma_{\mathrm{HI}}$ for only points \emph{in the ionized regions}, while the blue dashed curve is for the case where the effect of shadows is turned off.}
  \label{fig:various_one_z}
\end{figure*}

\begin{figure}
  \includegraphics[width=0.49\textwidth]{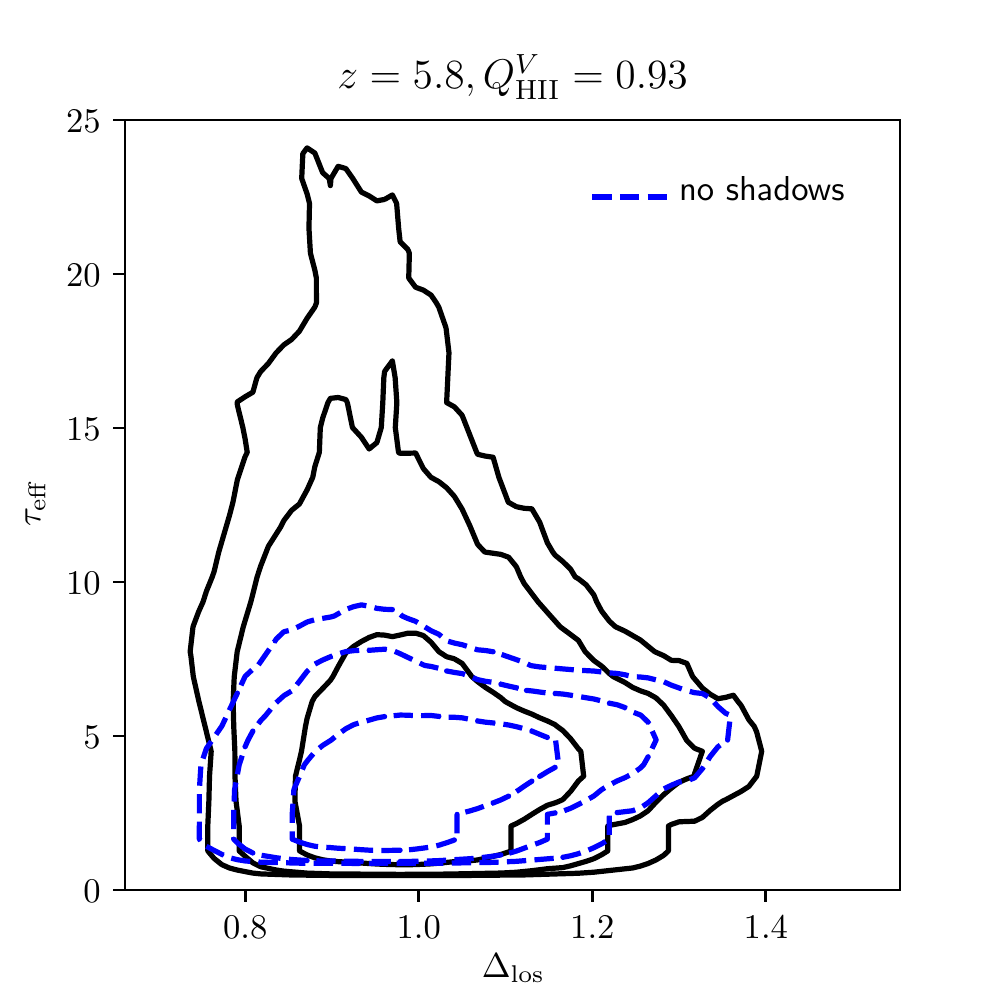}
  \caption{
The black curves show the distribution of sight lines in the $\tau_{\mathrm{eff}}- \Delta_{\mathrm{los}}$ plane, where $\tau_{\mathrm{eff}}$ is the effective optical depth and $\Delta_{\mathrm{los}}$ is the average overdensity along each line of sight. The distribution is calculated for the best-fit model obtained using the pessimistic data at $z = 5.8$. The contours contain $68\%$, $95\%$ and $99\%$ of the sight lines. For comparison, we also show the results when the shadows are turned off (blue dashed contours).
}
  \label{fig:taueff_Delta_one_z}
\end{figure}

Having obtained the allowed ranges in reionization history from our model, we now examine in some detail the properties of the best-fit model (red squares and black circles in \fig{fig:Qrange_MCMC}). We first show the CDF $P(<\tau_{\mathrm{eff}})$ for the best-fit model and the match with the data in \fig{fig:taueff}. In each panel, the gray shaded regions denote the predictions of our model accounting for statistical fluctuations along different sight lines while the red curves denote $P(<\tau_{\mathrm{eff}})$ for the  optimistic (left/upper) and pessimistic (right/lower) cases. The legend specifies which of the two cases is used for the model comparison.

The top panels show the predictions for the pessimistic data set at $z = 5$ (left), $5.2$ (middle), and $5.4$ (right), while the middle panels are at $z = 5.6$ (left), $5.8$ (middle), and $6$ (right). The bottom panels show the match for the optimistic data set at $z = 5.6$ (left), $5.8$ (middle) and $6$ (right). Since the best-fit models for the optimistic cases at $z = 5$, $5.2$ and $5.4$ are almost identical to the pessimistic cases (see Table \ref{tab:best-fit}), we do not show them separately in the figure. Visual comparisons between model and data can sometimes be misleading when the data points are highly correlated (which is the case here as different $\tau_{\mathrm{eff}}$ bins are indeed correlated), however, at least in this case, one can see that the best-fit model is a good description of the data. In particular, the model is able to reproduce the high values of $\tau_{\mathrm{eff}}$ that are required to match the data.  The data at $z \leq 5.4$ is consistent with a completely ionized medium and hence a relatively uniform $\Gamma_{\mathrm{HI}}$, which was also noted by \citet{Becker2015,Bosman2018}. However, at higher redshifts, the best-fit model contains some neutral patches in the IGM (characterized by a ionized volume fraction $Q^V_{\mathrm{HII}} < 1$). The introduction of neutral islands in the model allows for sight lines with large $\tau_\mathrm{eff}$, which captures the high-$\tau_\mathrm{eff}$ tail of the distribution.

Let us now understand the characteristics of the ionization and the radiation fields for our model. We choose the best-fit model corresponding to the pessimistic data set at $z = 5.8$ for which the effect of patchy reionization is quite prominent. The results are shown in \fig{fig:various_one_z}. The left hand panel shows the neutral fraction $x_{\mathrm{HI}}$ map for a two-dimensional slice of thickness $4 h^{-1}$~cMpc, while the middle panel shows the fluctuations in the photoionization rate $\Gamma_{\mathrm{HI}}$. The ionization field resembles an almost ionized universe with patches of neutral islands in between. As expected, $\Gamma_{\mathrm{HI}}$ is non-zero in the ionized regions and (almost close to) zero in the neutral regions, thus tracing the overall topology of the ionization map. What is interesting to note is that even in the ionized regions there are pixels close to the neutral islands where the photoionization rate is quite small. These regions are the ``shadows'' which arise because they do not receive photons from sources beyond the neutral islands. In our model, this effect is captured via the inclusion of the shadowing algorithm outlined in \secn{sec:photoionization}.

To see the effect of these shadows on the distribution of $\Gamma_{\mathrm{HI}}$ \emph{in the ionized regions}, we plot the PDF $\de P / \de \ln \Gamma_{\mathrm{HI}}$ (computed using only pixels in the ionized regions) in the right hand panel of \fig{fig:various_one_z} (the black line). One can see that the distribution peaks around the mean $\langle \Gamma_{\mathrm{HI}} \rangle$ and there exists a long tail for small values of $\Gamma_{\mathrm{HI}}$. This tail arises from the shadows near the neutral islands. To confirm this point, we have plotted in blue dashed curve the $\Gamma_{\mathrm{HI}}$ distribution for the case where we turn off the shadows, i.e., we simply use \eqn{eq:Gamma_HI_i_shells} instead of \eqn{eq:Gamma_HI_i_shadow} to compute the photoionization rate. Clearly, the absence of the shadows leads to a very sharply peaked distribution of $\Gamma_{\mathrm{HI}}$. Our $\Gamma_{\mathrm{HI}}$ distribution can be compared with those of \citet{Davies2016,Nasir2019} who also find a low-$\Gamma_{\mathrm{HI}}$ tail. In fact, \citet{Nasir2019} have explicitly checked that such a tail arises from shadowing of the neutral islands (see their Fig. 5). In this sense, our findings are qualitatively similar to theirs, although the implementations are very different. 

We further compare our model with that of \citet{Davies2016} by computing the mean overdensity $\Delta_{\mathrm{los}}$ and the mean effective optical depth $\tau_{\mathrm{eff}}$ along different sight lines of length $50 h^{-1}$~cMpc and plotting the resulting two-dimensional distribution in \fig{fig:taueff_Delta_one_z}. The black contours correspond to the distribution for the best-fit model, while the blue dashed contours are for the case where the shadows have been tuned off. The three contours in each case enclose $68\%$, $95\%$, and $99\%$ of sight lines from the sample. It is clear that there is a mild negative correlation between $\Delta_{\mathrm{los}}$ and $\tau_{\mathrm{eff}}$ for the best-fit model, entirely different from the positive correlation that is expected for an IGM with uniform $\Gamma_{\mathrm{HI}}$, see \eqn{eq:tau_alpha_Gamma_HI}. Our model is thus qualitatively consistent with \citet{Davies2016}. We further conclude from the figure that the presence of shadows increases the fluctuations in $\tau_{\mathrm{eff}}$ allowing for a better fit to the data for the pessimistic case.

\section{Summary and discussion}
\label{sec:discussion}

Recent observations of the effective optical depth $\tau_{\mathrm{eff}}$ of Ly$\alpha$ absorption at $5 \lesssim z \lesssim 6$ show significant fluctuations when averaged over reasonably large scales $50 h^{-1}$~cMpc scales. One possible interpretation of these observations is that the fluctuations arise because of left-over HI islands and that HI reionization is complete only at $z \sim 5.2$ \citep{Kulkarni2019,Keating2019,Nasir2019}. If this interpretation of the data is indeed true, it becomes imperative to include these observations  in any parameter constraints related to the reionization history (in addition to, e.g., the existing CMB observations of electron scattering optical depth). Obtaining constraints, in turn, requires efficient methods of computing the relevant observables (in this case, the Ly$\alpha$ optical depth) to probe the space of unknown parameters.

To achieve this goal, we have developed a semi-numerical technique to constrain the reionization history at $5 \lesssim z \lesssim 6$. Our method is appropriate for probing large-scale properties of the Ly$\alpha$ absorption in relatively low-resolution simulation boxes and relies on two main inputs: (i) the modelling of ionized regions using a photon-conserving semi-numerical code of reionization  \citep[\texttt{SCRIPT};][]{Choudhury2018} combined with a prescription for blocking photons from sources along sight lines passing through neutral regions and (ii) modelling the Ly$\alpha$ optical depth using the fluctuating Gunn-Peterson approximation. To our knowledge, this is the least computationally expensive model to study the Ly$\alpha$ opacity fluctuations.

We find that the model is able to capture the essential properties of the HI field as observed in the Ly$\alpha$ absorption, similar to those found in other semi-numerical models \citep{Davies2016,Nasir2019} and more detailed simulations \citep{Kulkarni2019,Keating2019}. Since the method is computationally fast, it allows us to probe the parameter space quite efficiently and thus obtain the range of histories consistent with the data (keeping in mind that other physical quantities like the temperature and the amplitude of the photoionization rate cannot be constrained by our model). We find that the inferred reionization history is delayed when we use the data set where non-detections of the flux are treated as having infinite optical depth \citep[the so-called `pessimistic' case of][]{Bosman2018} compared to the case where non-detections are assumed to have optical depths just below the detection limit (the `optimistic' case). The data are \emph{in}consistent with reionization being complete at $z > 5.6$ (independent of which data set is used). The completion can be as late as $z \sim 5.2$, corresponding to the $2\sigma$ lower limits on the ionized fraction. We also find that the ionized volume fraction can be as low as $\sim 60\%$ at $z \sim 6$ for the pessimistic data set. The analysis thus indicates the potential of our technique in constraining the reionization history with more number of quasar sight lines at $z\sim 6$.

The number of known $z>5$ quasars will increase dramatically in the next decade with the upcoming quasar searches which will be performed by EUCLID \citep{Barnett2019, Griffin2020}, the Vera Rubin Observatory \citep[formerly LSST;][]{LSST} and the Nancy Grace Roman Space Telescope (formerly WFIRST; \citealt{NGR}). Coupled with more efficient spectroscopic observations owing to the ELT \citep{Gilmozzi2007}, new high-$z$ quasars will significantly increase the amount and quality of Ly$\alpha$ opacity information at $z>6$ in the next decade. The discovery of bright quasars beyond $z=7.5$ \citep{Banados2018} ensures we will be able to map the large-scale evolution of opacity until $z\sim6.5$, where Gunn-Peterson absorption is expected to saturate fully.

Indeed, fast semi-numerical models like ours often are unable to track all the physical processes self-consistently. We have also seen in \fig{fig:Qrange_MCMC} that the reionization history inferred from our analysis can be different from that predicted by other simulations \citep{Kulkarni2019,Keating2019,Nasir2019}. One possible source of uncertainty in our model arises from the treatment of the regions which do not receive photons from sources whose lines of sight pass through the HI islands, thus creating shadows and suppressing the photoionization rate. Another possibility could be that our assumed value of mean free path $\lambda_{\mathrm{ss}}$ is different than that in reality. In the future, we plan to make a detailed comparison of our model with simulations to check the validity of our method of producing shadows as well as test various other assumptions of the model.

We also plan to further improve the analysis by including various effects which have been ignored in this work. The first is to model the Ly$\beta$ absorption from the same sight lines and compare with the available data \citep{Eilers2019}, thus obtaining more stringent constraints on the reionization history. The second is to forward model the noise in the observations and include it in the model, instead of considering the two extreme cases as done in this work. The third and perhaps most important improvement would be to self-consistently model the temperature evolution in each cell across redshifts. This would relax the assumption of the power law relation between the temperature and density of the cell and should be able to account for the dependence of the temperature field on the reionization history. In addition, we also plan to expand the scope of the analysis to include other data sets and hence constrain the reionization history accounting for causal correlations between different redshift bins. Such improvements, combined with the fact that the model is computationally inexpensive, would then allow for comparing with a wide variety of observations simultaneously and hence obtain constraints on reionization.

\section*{Acknowledgements}
TRC acknowledges support of the Department of Atomic Energy, Government of India, under project no.~12-R\&D-TFR-5.02-0700 and the Associateship Scheme of ICTP. The research of AP is supported by the Associateship Scheme of ICTP, Trieste and the Ramanujan Fellowship awarded by the Department of Science and Technology, Government of India. SEIB acknowledges funding from the European Research Council (ERC) under the European Union’s Horizon 2020 research and innovation program (grant agreement No.~669253). We thank the anonymous referee for constructive comments which helped improve the content of the paper.

\section*{Data availability}
The code for generating the ionized regions, \texttt{SCRIPT}, is available at \texttt{https://bitbucket.org/rctirthankar/script}. The rest of the data underlying this article will be shared on reasonable request to the authors.

\bibliographystyle{mnras}
\bibliography{lya_fluctuations}

\appendix

\section{Dependence of the results on the parameter choices}
\label{app:parameters}

\begin{figure*}
  \includegraphics[width=0.9\textwidth]{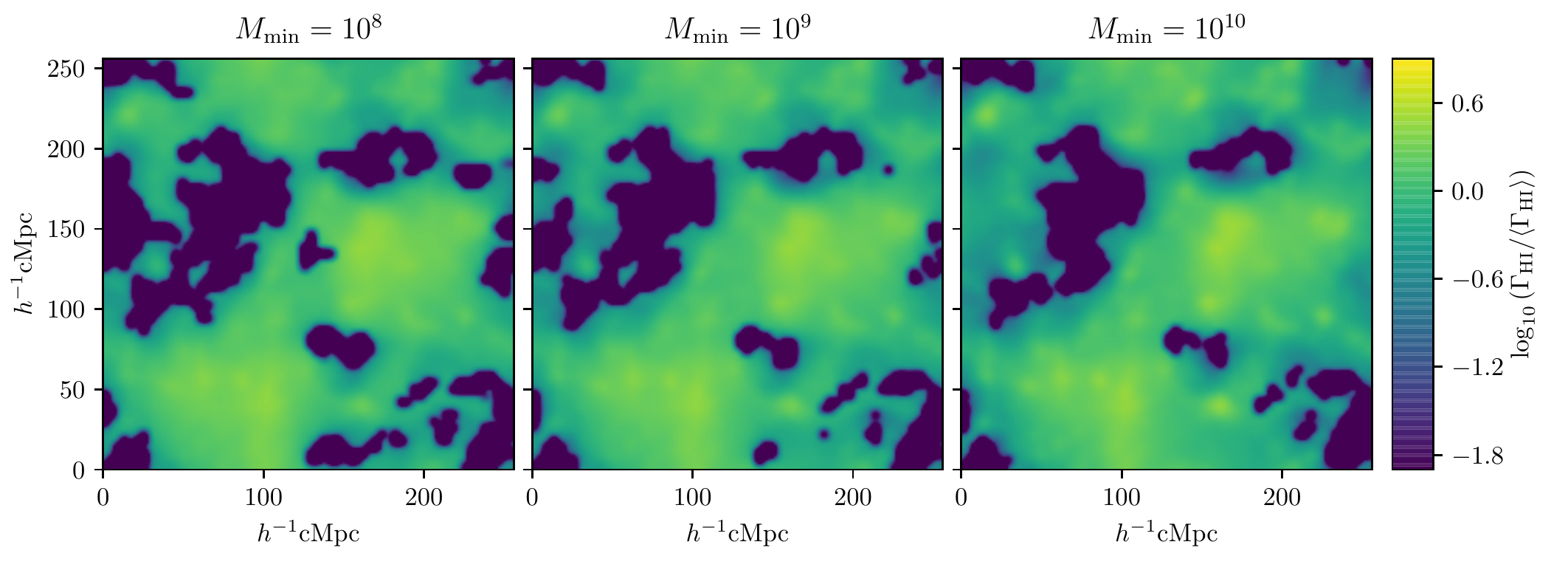}
  \caption{
The two-dimensional maps of the photoionization rate fluctuations for a slice of thickness $4 h^{-1}$~cMpc at $z = 5.8$ for different values of $M_{\mathrm{min}}$ (in units of $\Msun$) as mentioned above the respective panels.
}
  \label{fig:Gamma_map_Mmin_onez}
\end{figure*}

\begin{figure}
  \includegraphics[width=0.47\textwidth]{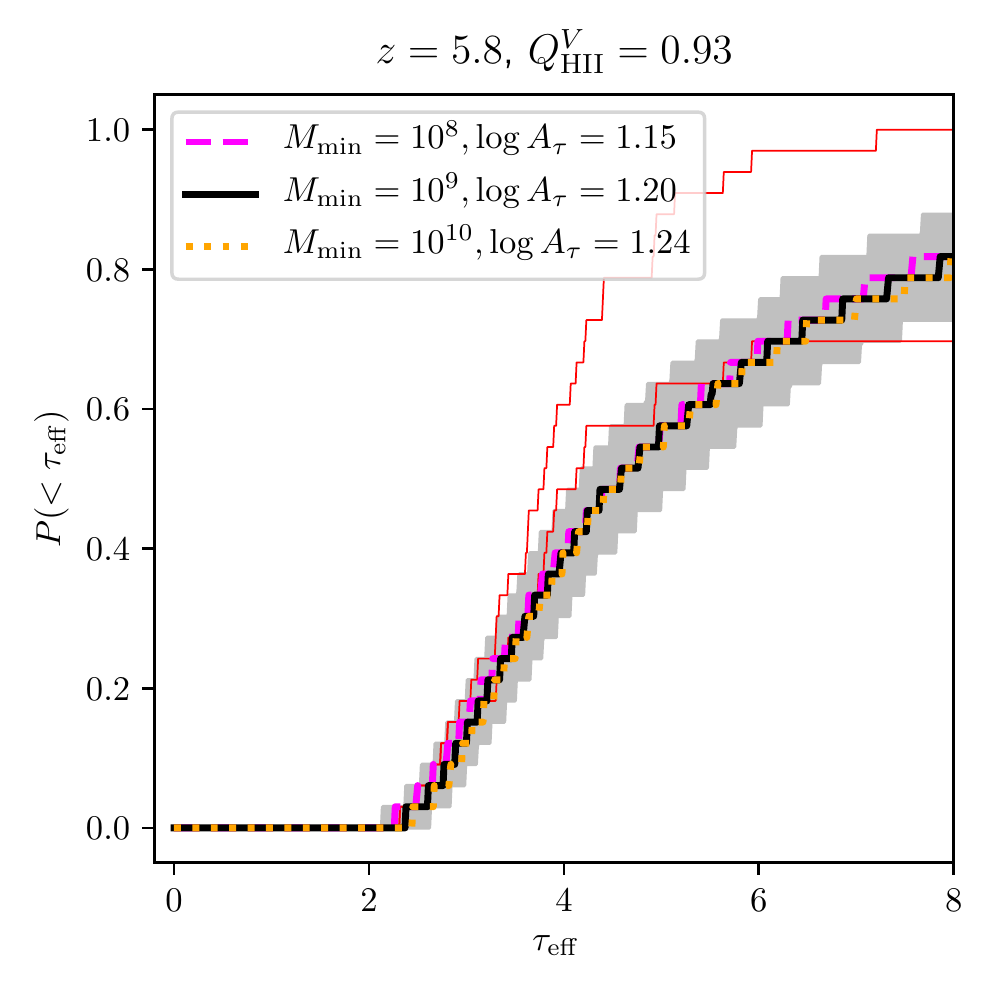}
  \caption{
The $\tau_{\mathrm{eff}}$ CDF for three different values of $M_{\mathrm{min}}$ at $z = 5.8$ (with red curves showing the optimistic and pessimistic data sets used in the paper). The default runs in the paper are for $M_{\mathrm{min}} = 10^9 \Msun$. All the three cases have the same value of $Q^V_{\mathrm{HII}}$ (corresponding to the best-fit value in the default case), while the $A_{\tau}$ values are adjusted so as to match the mean $\tau_{\mathrm{eff}}$ for the default case. All other model parameters are identical for the three cases.
}
  \label{fig:Mmin_onez}
\end{figure}

\begin{figure}
  \includegraphics[width=0.51\textwidth]{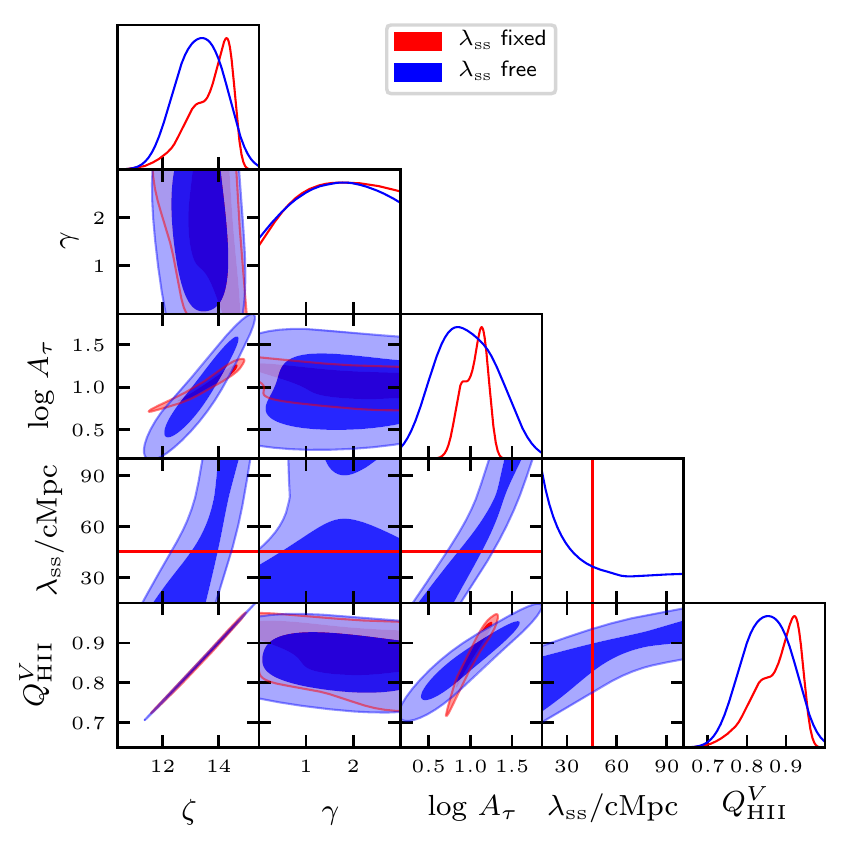}
  \caption{ 
The posterior distribution of parameters obtained using MCMC analysis when $\lambda_{\mathrm{ss}}$ is kept fixed to the default value (red) and when $\lambda_{\mathrm{ss}}$ is left free (blue). The results are shown for the pessimistic data set at $z = 5.8$. For the case where $\lambda_{\mathrm{ss}}$ is fixed, the red straight lines parallel to the axes denote the fixed value of $\lambda_{\mathrm{ss}}$.
  }
  \label{fig:getdist_mfp_triangle}
\end{figure}

\begin{figure*}
  \includegraphics[width=0.9\textwidth]{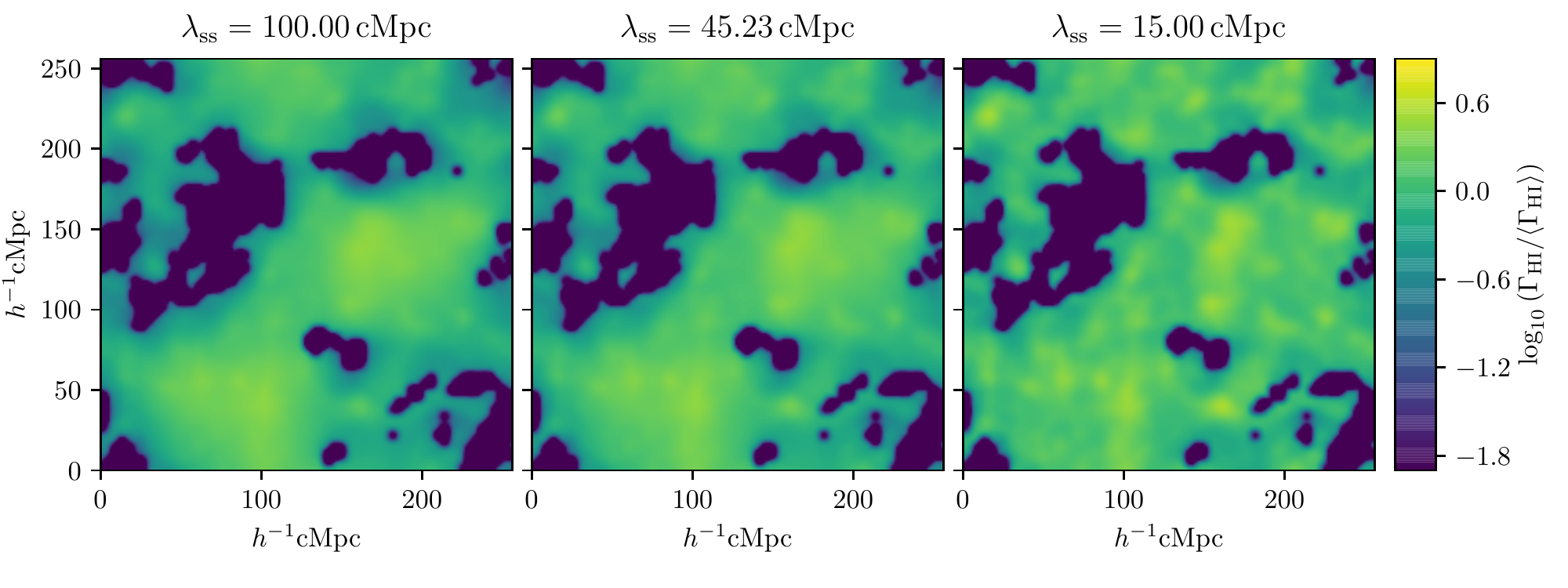}
  \caption{Same as \fig{fig:Gamma_map_Mmin_onez} but for different values of $\lambda_{\mathrm{ss}}$.}
  \label{fig:Gamma_map_mfp_onez}
\end{figure*}

\begin{figure}
  \includegraphics[width=0.47\textwidth]{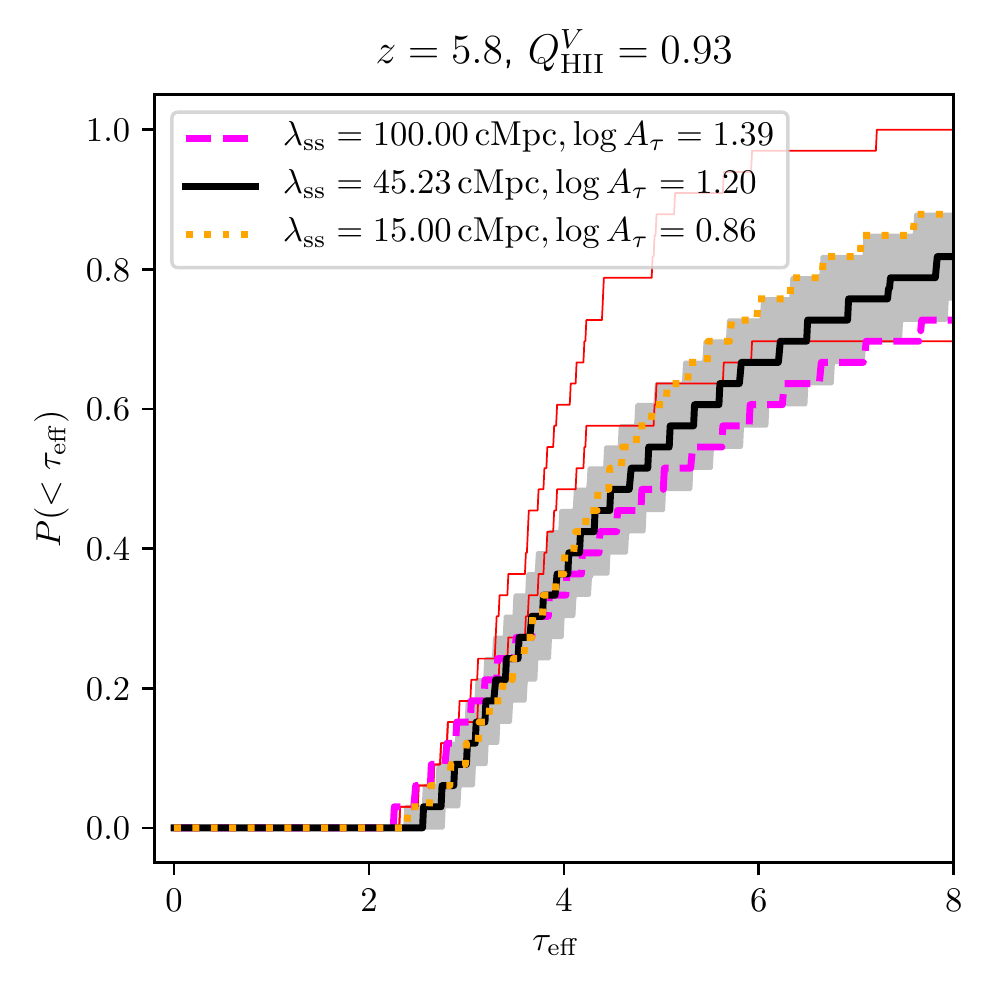}
  \caption{Same as \fig{fig:Mmin_onez} but for different values of $\lambda_{\mathrm{ss}}$.}
  \label{fig:mfp_onez}
\end{figure}

In this appendix, we discuss the dependence of our model on different parameters once we vary them beyond their default values.

\subsection{The minimum halo mass $M_{\mathrm{min}}$}
\label{app:Mmin}

We have fixed the value of the minimum mass $M_{\mathrm{min}}$ of haloes that can produce ionizing photons to $10^9 \Msun$ for the default analysis in \secn{sec:constraints}. In this section, we investigate if the choice affects our constraints on the reionization history. Let us first see the $\Gamma_{\mathrm{HI}}$ fluctuations for a two-dimensional slice for three values of $M_{\mathrm{min}}$ at a representative redshift, chosen to be $z = 5.8$. The maps are shown in \fig{fig:Gamma_map_Mmin_onez}. We have fixed the value of $Q^V_{\mathrm{HII}}$ for the three cases (by changing the $\zeta$ values appropriately). We can see that although the large-scale properties look very similar, there are some obvious differences. As $M_{\mathrm{min}}$ increases, the neutral regions (shown by the dark points, corresponding to $\Gamma_{\mathrm{HI}}$ values almost zero), which are relatively smaller in size, tend to disappear. Since the amount of neutral volume is the same, the existing neutral regions have larger sizes for higher $M_{\mathrm{min}}$.

However, these changes, which are visually obvious, seem to affect the maps at scales smaller than $50 h^{-1}$~cMpc. As a result, we expect that these may not affect the $\tau_{\mathrm{eff}}$ distributions when averaged over large path lengths. To confirm this, we plot the CDF $P(<\tau_{\mathrm{eff}})$ for the three cases in \fig{fig:Mmin_onez}. For the default case $M_{\mathrm{min}} = 10^9 \Msun$, we choose parameters corresponding to the best-fit model obtained in \secn{sec:constraints}. For the other two cases, we keep the same value of $Q^V_{\mathrm{HII}}$ and adjust $A_{\tau}$ to match the mean $\tau_{\mathrm{eff}}$ for the default case. As is clear from the figure, the three models produce identical $P(<\tau_{\mathrm{eff}})$, thus confirming our expectation that the $Q^V_{\mathrm{HII}}$ constraints are insensitive to $M_{\mathrm{min}}$ (although the values of $A_{\tau}$ could vary by $\sim 10\%$ as we change $M_{\mathrm{min}}$ by an order of magnitude). It is possible that models with different $M_{\mathrm{min}}$ can be distinguished when the transmitted flux is averaged over sight lines of smaller length; this is a possible avenue which we plan to explore in the future.

We did not investigate whether the $\tau_{\mathrm{eff}}$ distributions change more substantially for a reionization model dominated by much rarer sources, e.g., AGNs \citep{Kulkarni2017}. Such models would require higher values of $M_{\mathrm{min}}$ than what is considered here. In such cases, our sub-grid based method of computing the collapsed fraction becomes less accurate and hence a more careful analysis is warranted. We plan to study such cases separately in the future.

\subsection{The mean free path $\lambda_{\mathrm{ss}}$}
\label{app:mfp}

The default value of the mean free path $\lambda_{\mathrm{ss}}$, determined by the distance between the self-shielded regions, has been chosen as the value extrapolated from lower redshift observations. In this section, we investigate if the uncertainty in the value of $\lambda_{\mathrm{ss}}$ affects our constraints on $Q^V_{\mathrm{HII}}$ as discussed in \secn{sec:constraints}.

We first perform a full MCMC analysis treating $\lambda_{\mathrm{ss}}$ as a free parameter. We choose it to have a flat prior in the range $15$ to $100$~cMpc. These limits are set by the resolution and the box size of our simulation, however, they cover the most interesting range of values for the parameter. The posterior distribution of the parameters for this case, along with those of our three-parameter default run with $\lambda_{\mathrm{ss}}$ fixed, are shown in \fig{fig:getdist_mfp_triangle}. The constraints are obtained using the pessimistic data set at $z = 5.8$. We can immediately see that the posterior distribution of $Q^V_{\mathrm{HII}}$ is different when $\lambda_{\mathrm{ss}}$ is left free. The best-fit value is smaller than the default case. However, interestingly, the $2\sigma$ limits on $Q^V_{\mathrm{HII}}$ remain relatively unaffected (the range is $[0.76-0.96]$ for the default case, while it is modified to $[0.73-0.97]$ when $\lambda_{\mathrm{ss}}$ is free). Hence we can conclude that our constraints on the reionization history remain relatively unaffected even when $\lambda_{\mathrm{ss}}$ is allowed to vary.

We find that it is \emph{not} possible to constrain $\lambda_{\mathrm{ss}}$ from our analysis. The marginalized posterior is almost flat with marginal increase at lower values. In our approach, it is perhaps more convenient to make use of constraints on $\lambda_{\mathrm{ss}}$ from other studies, either observations or simulations.

We can see from the figure that there is a positive correlation between $\lambda_{\mathrm{ss}}$ and $\log A_{\tau}$. The reason is that a larger mean free path allows sources from larger distances to contribute to the photon flux $S_i$, thus decreasing the optical depth $\tau_{\alpha, i}$. One requires a higher value of $A_{\tau}$ to compensate for this increase in the flux and increase $\tau_{\alpha, i}$, see \eqn{eq:tau_alpha}. This correlation between $\lambda_{\mathrm{ss}}$ and $\log A_{\tau}$ leads to a wider range of $A_{\tau}$ values to be allowed than that in the default case.

There is also a positive correlation between $\lambda_{\mathrm{ss}}$ and $Q^V_{\mathrm{HII}}$ (or $\zeta$). To understand this, we compare the $\Gamma_{\mathrm{HI}}$ maps for different values of $\lambda_{\mathrm{ss}}$ in \fig{fig:Gamma_map_mfp_onez}. For this plot, we have fixed the value of $Q^V_{\mathrm{HII}} = 0.93$ for all the three cases. In the left hand panel, we choose a high value $\lambda_{\mathrm{ss}} = 100$~cMpc, which essentially implies that points in the ionized regions can see sources at large distances (unless blocked by a neutral region). Thus, for large $\lambda_{\mathrm{ss}}$, only points that are considerably far away from the islands (say, with distance $\gtrsim \lambda_{\mathrm{ss}}$) can receive photons from all directions without being obstructed by the islands. This implies that the number of points that are affected by these islands is relatively larger, which then leads to more fluctuations. Hence, we end up with a somewhat counter-intuitive result where increasing $\lambda_{\mathrm{ss}}$ leads to more large-scale fluctuations in $\Gamma_{\mathrm{HI}}$. It follows that, to produce the same amount of fluctuations as the default $\lambda_{\mathrm{ss}}$, we require less neutral islands and thus larger $Q^V_{\mathrm{HII}}$ for the case of a higher $\lambda_{\mathrm{ss}}$. This is the cause of the positive correlation found in the MCMC analysis.

In the right panel of \fig{fig:Gamma_map_mfp_onez}, we show the $\Gamma_{\mathrm{HI}}$ map for a smaller value $\lambda_{\mathrm{ss}} = 15$~cMpc. As expected, we find a lot of small-scale fluctuations in $\Gamma_{\mathrm{HI}}$ in this case. Moreover, only points that are very close to the neutral islands ($\lesssim 15$~cMpc) are affected by them, leading to a much smaller effect of the shadows.

The effect of $\lambda_{\mathrm{ss}}$ on the $\tau_{\mathrm{eff}}$ CDF is shown in \fig{fig:mfp_onez}. Note that we have kept the value of $Q^V_{\mathrm{HII}} = 0.93$ same for the three cases plotted, while the value of $A_{\tau}$ is chosen to obtain the same mean $\tau_{\mathrm{eff}}$. As expected, the $\tau_{\mathrm{eff}}$ CDF is wider for larger $\lambda_{\mathrm{ss}}$ because of more large-scale fluctuations, and similarly narrower for smaller $\lambda_{\mathrm{ss}}$. However, the variation even in such extreme cases is within the cosmic variance of the observable, thus not affecting the constraints on $Q^V_{\mathrm{HII}}$ that severely.

To summarize the results of this section, we find that the constraints on the reionization history are, in principle, affected by the chosen value of $\lambda_{\mathrm{ss}}$. In addition, we also find that our model is unable to put any constraints on $\lambda_{\mathrm{ss}}$.

\section{Convergence of the results with respect to resolution}
\label{app:resolution}

\begin{figure}
  \includegraphics[width=0.49\textwidth]{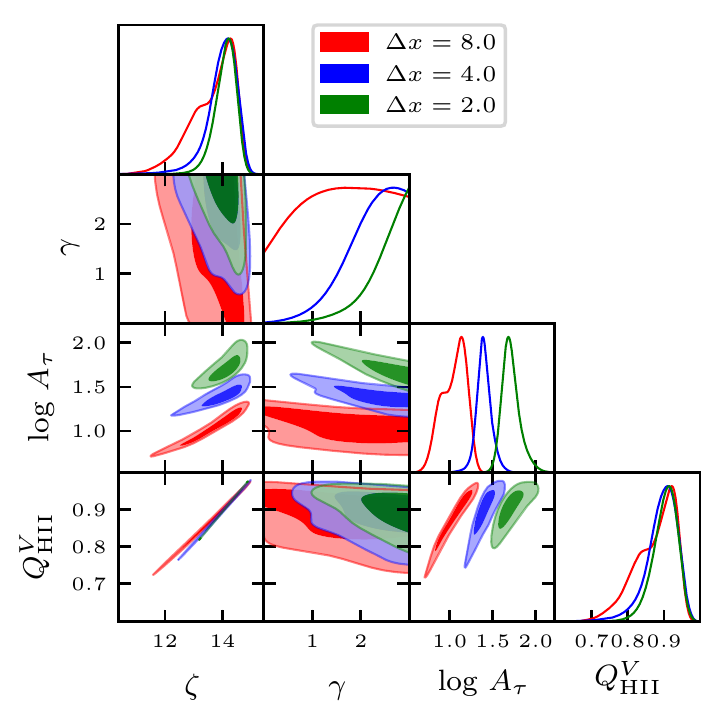}
  \caption{
The posterior distribution of parameters obtained using MCMC analysis for three different grid resolutions used for generating the ionization maps and the Ly$\alpha$ optical depth. The results are shown at $z = 5.8$ for the pessimistic data set. The default run in the paper corresponds to $\Delta x = 8 h^{-1}$~cMpc (red) whose results are identical to those shown by red contours/curves in \fig{fig:getdist_onez_triangle}.
  }
  \label{fig:getdist_resolution_triangle}
\end{figure}

\begin{figure}
  \includegraphics[width=0.47\textwidth]{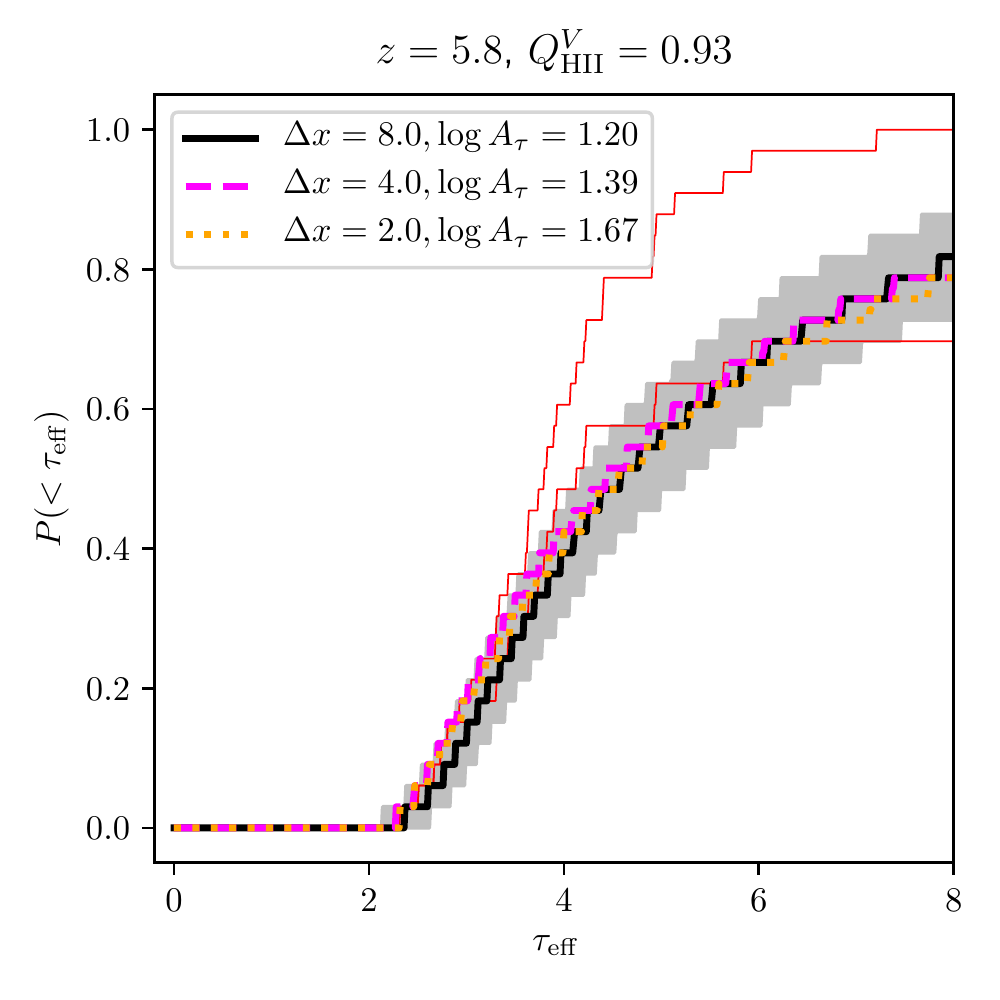}
  \caption{
The $\tau_{\mathrm{eff}}$ CDF for three different grid resolutions used while generating the HI field and the Ly$\alpha$ optical depth at $z = 5.8$ (with red curves showing the optimistic and pessimistic data sets used in the paper). The model parameters correspond to the best-fit values found from the MCMC run shown in \fig{fig:getdist_resolution_triangle}. Note that all the three cases have the same value of $Q^V_{\mathrm{HII}}$, while the $A_{\tau}$ values are mentioned in the legend.
}
  \label{fig:resolution_onez}
\end{figure}

In this appendix, we study the dependence of our results on the grid size chosen for generating the ionized bubbles and computing the Ly$\alpha$ optical depth. This is particularly important as some of the model parameters depend on the resolution. For this study, we have chosen the pessimistic data set at $z=5.8$ and carried out the MCMC analysis for grid sizes of $\Delta x = 4 h^{-1}$~cMpc ($64^3$ grids) and $\Delta x = 2 h^{-1}$~cMpc ($128^3$ grids). Recall that our default runs are for a coarser resolution $\Delta x = 8 h^{-1}$~cMpc ($32^3$ grids). While running for the different resolutions, we choose the same priors as the default case and fix $M_{\mathrm{min}}$ and $\lambda_{\mathrm{ss}}$ to their default values as mentioned in \secn{sec:free_params}. The posterior distributions of the various parameters for the three cases are shown in \fig{fig:getdist_resolution_triangle}.

The first point to note from the figure is that the allowed values of $A_{\tau}$ decrease with coarsening resolution (i.e., increasing $\Delta x$). This is not surprising as $A_{\tau}$ is directly dependent of $\kappa_{\mathrm{res}}$, a parameter that itself depends on the resolution. The fact that the constraints on $A_{\tau}$ are resolution-dependent is a direct consequence of using the fluctuation Gunn-Peterson approximation and our inability to capture the small-scale fluctuations in the low resolution simulations. We also find that the other parameter which is affected by resolution, namely, $\gamma$, is also different for the different resolutions. However, the constraints on $\gamma$ are weak no matter which resolution is chosen and $\gamma \approx 1.5-2$ seems to be a good choice for all the cases.

What is important for our analysis is that the constraints of $Q^V_{\mathrm{HII}}$ remain similar for the three resolutions. The best-fit values are almost the same (they differ by only $\lesssim 0.01$). The $2\sigma$ upper limits are also within $\sim 0.005$ of each other. There is some difference in the three cases at the lower tail of the $Q^V_{\mathrm{HII}}$ posterior distribution. The $2\sigma$ lower limit for the default case $\Delta x = 8 h^{-1}$~cMpc is $0.76$, while they are $0.80$ and $0.82$ for $\Delta x = 4 h^{-1}$~cMpc and $\Delta x = 2 h^{-1}$~cMpc respectively. This implies that our default runs underestimate the lower $2\sigma$ values by $\sim 0.06$ compared to the highest resolution probed here. The reason for $Q^V_{\mathrm{HII}}$ lower limits to be more stringent for higher resolution is as follows: more neutral IGM leads to more fluctuations in $\tau_{\mathrm{eff}}$, whereas coarser resolution simulations tend to smooth the fluctuations. Hence, coarser resolution simulations allow agreement with the data even for lower values of $Q^V_{\mathrm{HII}}$ (values which produce more fluctuations in the finer resolution runs and thus tend to get ruled out). The resolution-dependence of the constraints is expected to be less prominent for cases where the allowed $Q^V_{\mathrm{HII}}$ values are higher, e.g., for the optimistic data sets and for lower redshifts.

We choose $\Delta x = 8 h^{-1}$~cMpc as our default grid size because the MCMC runs take much less time to complete. For the three cases plotted in \fig{fig:getdist_resolution_triangle}, it takes $\sim 6$ hours, $\sim 36$ hours, and $\sim 15$ days for the runs to complete for $\Delta x = 8 h^{-1}$~cMpc, $4 h^{-1}$~cMpc and $2 h^{-1}$~cMpc respectively. Hence, using a coarse resolution allows us to perform many more MCMC runs in a reasonable amount of time and study the different features of the model in more detail. The downside of using the coarse resolution is that the lower limits quoted are somewhat conservative.

For completeness, we also show the CDF $P(<\tau_{\mathrm{eff}})$ for the three resolutions in \fig{fig:resolution_onez}. We choose the best-fit $Q^V_{\mathrm{HII}}$ and $A_{\tau}$ values for the three resolutions as found from the MCMC runs and fix $\gamma = 1.5$ for all the cases. It is clear that the models produce almost identical $\tau_{\mathrm{eff}}$ distributions for the same value of $Q^V_{\mathrm{HII}}$ although the $A_{\tau}$ values are different. This is consistent with our findings that the reionization constraints are insensitive to the grid size.

Overall, we can safely conclude that our results are not sensitive to the resolution, except for the lower limits being slightly underestimated for the coarser resolution. We use $\Delta x = 8 h^{-1}$~cMpc for the MCMC analysis in the paper.

We have also checked whether our results are sensitive to the box size by comparing with a smaller box of length $128 h^{-1}$~cMpc. We find that the $\tau_{\mathrm{eff}}$ distributions remain unchanged for the smaller box. This is perhaps not surprising as all the relevant scales in the problem (the size of the neutral regions, the mean free path, and the length of the sight lines over which the optical depth is calculated) are smaller than the boxes used.

\end{document}